# Dynamics of Small Autocatalytic Reaction Networks

# II : Replication, Mutation and Catalysis

By

PETER F. STADLER[a,b], WOLFGANG SCHNABL[a,†],

CHRISTIAN V. FORST[c], AND PETER SCHUSTER[a,b,c,⋆]

[a] Institut für Theoretische Chemie der Universität Wien

[b] The Santa Fe Institute, Santa Fe, New Mexico

[c] Institut für Molekulare Biotechnologie e.V., Jena

‡ Present Address: Alcatel Austria Research Center, Wien

⋆ Address Correspondence to:

Professor Peter Schuster, Institut für Molekulare Biotechnologie e.V.

Beutenbergstraße 11, PF 100 813, D-07708 Jena, Germany

Phone: **49 (3641) 65 6333

Fax: **49 (3641) 65 6335

E-Mail: pks @ imb-jena.de

**Abstract.**

Mutation is introduced into autocatalytic reaction networks. The differential equations obtained are neither of replicator type nor can they be transformed straightway into a linear equation. Examples of low dimensional dynamical systems — $n = 2, 3$ and $4$ — are discussed and complete qualitative analysis is presented. Error thresholds known from simple replication-mutation kinetics with frequency independent replication rates occur here as well. Instead of cooperative transitions or higher order phase transitions the thresholds appear here as supercritical or subcritical bifurcations being analogous to first order phase transitions.

## 1. Introduction

About ten years ago Schuster and Sigmund (1983) pointed out that the same class of model equations applies to the description of population dynamics in different disciplines ranging from chemical kinetics to theoretical sociobiology and they coined the terms **replicator dynamics** for the common phenomena and **replicator equations** for the corresponding ODEs, respectively. These simple models deal exclusively with (error-free) replication in populations of constant size. We are concerned here with more complex dynamics which allows for mutations within a set of variants, all of which are considered explicitly as variables. In particular we study some low-dimensional examples for which analytical treatments are possible, and present a few results for arbitrary dimensions. Low-dimensional replicator equations with reaction channels for error copies are for example relevant models for introducing mutation into

- chemical kinetics of catalyzed replication,
- Fisher's selection equation,
- dynamics of coevolution, and
- dynamics of Maynard Smith games.

It is worth mentioning that replicator equations of dimension $n$ are dynamically equivalent to Lotka-Volterra equations of dimension $n-1$ (Hofbauer, 1981).

Explicit consideration of mutation terms is highly relevant for realistic models since they describe processes of fundamental importance in nature which are often neglected only because they make the handling of the model equations much harder, and they are commonly prohibitive for analytical solutions. An often useful compromise is shortly mentioned here: variants are not described by explicit variables but rather lumped together into a so-called "error-tail" as suggested in a stochastic analysis of error propagation by Nowak and Schuster (1989). This simplification allows to treat cases with more elaborate dynamics and autocatalytic reaction networks (Andrade *et al.*, 1993; Nuño *et al.*, 1993a,b; Stadler and Nuño, 1994).

Replication processes, correct and incorrect, are fundamental to selection and evolutionary dynamics in general. In the simplest case which refers to asexual reproduction in constant environments replication rates are proportional to template concentrations and the replication rate parameters are constants. Then populations approach a unique asymptotically stable steady state which has been characterized as *(molecular) quasi-species* since these stationary mutant distributions represent the genetic reservoirs of asexually replicating species as much as the *gene pools* do in ordinary biological species. The formal mathematical background of the quasi-species concept was updated and its relation to experimental data obtained in test tube evolution experiments and in virology was summarized in a recent review (Eigen *et al.*, 1988, 1989).

Replication dynamics becomes more complex if the restriction to frequency independent replication rate constants is dismissed (as we do here). Frequency dependent replication functions are readily interpreted in terms of chemical reaction kinetics: positive sign means catalysis, negative sign implies inhibition. In the simplest case the catalyst is a molecule of the same class as the template. Examples are now well known in the biochemical literature: certain RNA molecules can act as specific catalysts for processing (Cech, 1986) or replication of other RNA molecules (Doudna and Szostak, 1989). Catalyzed replication is thus dealing with two forms of catalytic action (section 2): one RNA molecule is copied and acts as autocatalyst or template and a second one has the catalytic function (like a conventional protein enzyme in conventional biochemistry). Biochemical catalysis usually follows complex multistep mechanisms whose detailed dynamics is not accounted for by the single step kinetics of replicator equations. Replicator dynamics, however, provides straightforward insight into phenomena observed under some limiting conditions. As an example we mention "faster-than-exponential" growth predicted by the hypercycle equation (a special class of replicator equations discussed in section 2; see also Eigen 1971, Eigen and Schuster, 1979) as recently found experimentally with replication of RNA viruses in host cells (Eigen *et al.*, 1991).

Frequency dependent error-free replication has been studied extensively in the past (for a comprehensive survey see Hofbauer and Sigmund, 1988). All kinds of complex dynamical behaviour including chaotic attractors (Schnabl *et al.*, 1991) were observed. Introduction of mutation into these dynamical systems provides substantial difficulties for qualitative analysis (see section 2) and very few investigations with explicit and exact consideration of mutations were performed so far (for a general approach to mutation based on perturbation theory see Stadler and Schuster, 1992). There is one important exception: the selection mutation equation derived from Fisher's selection equation by explicit consideration of mutations. It leads to much simpler dynamical phenomena as oscillations and deterministic chaos can be excluded, and analytical results are available on mutations in this particular case (Hadeler, 1981; Hofbauer, 1985).

In this paper we shall derive exact results complemented by numerical studies on autocatalytic replication networks including explicit mutation terms. At first we introduce the replication-mutation equation in section 2 and consider the rest point in the central part of the physically meaningful domain of the phase space. In section 3 the two species model is presented in great detail. Section 4 and section 5 are dealing with mutation in special cases of replicator equations, in the hypercycle model (Eigen and Schuster, 1979) and in the multidimensional generalized Schlögl model (Schlögl, 1972; Schuster, 1986). In the latter case we had to restrict the analysis to low dimensions ($n = 3$ and $4$). A few results are available also for the limit of large dimensions ($\lim n \to \infty$).

## 2. The replication-mutation equation

*2.1. The model and its assumptions*

The fundamental process basic to the simplest class of autocatalytic reaction networks with frequency dependent replication rates is of the form

$$(\mathbf{A}) + \mathbf{I}_i + \mathbf{I}_j \xrightarrow{q_{ki} \cdot a_{ij}} \mathbf{I}_k + \mathbf{I}_i + \mathbf{I}_j ; \quad i,j,k = 1,\ldots,n . \tag{1}$$

**A** is the substrate needed for replication. Its concentration is assumed to be buffered and does not enter the kinetic equations as a variable (and is hence written in parentheses). The replicator produced in the reaction is the *target* $\mathbf{I}_k$, $\mathbf{I}_i$ is the template which is replicated, and $\mathbf{I}_j$ is the catalyst, all three belonging to the same class of species. The non-negative rate constants for these reactions are understood as elements of an $n \times n$ matrix $A \doteq (a_{ij} \geq 0)$. We assume that the species (the term species is used here for defined biochemical or chemical entities entering the kinetic equations as individual variables and not in the biological sense) $\mathbf{I}_i$ replicates correctly with the frequency $q_{ii}$. A mutation from $\mathbf{I}_i$ to $\mathbf{I}_k$, ($k \neq i$) occurs with the frequency $q_{ki}$. The product of a replication process has to be either a correct copy or a mutant and hence the conservation law

$$\sum_{k=1}^{n} q_{ki} = 1 \tag{2}$$

holds. Thus, the mutation matrix $Q \doteq (q_{ki})$ is a (column) stochastic matrix. Mutation frequencies depend on template $\mathbf{I}_i$ and target $\mathbf{I}_k$, but not on the catalyst $\mathbf{I}_j$. The relative concentration of a species is denoted by $[\mathbf{I}_k]/C = x_k$, where $C = \sum_{k=1}^{n}[\mathbf{I}_k]$, therefore $\sum_{k=1}^{n} x_k = 1$.

Application of mass action kinetics to the reaction-mutation network (1) under the constraint of constant total concentration $C$ yields the replication-mutation equation, an ODE of the form

$$\dot{x}_k = \sum_{i=1}^{n}\sum_{j=1}^{n} q_{ki} a_{ij} x_i x_j - x_k \Phi ; \quad k = 1,\ldots,n . \tag{3}$$

An unspecific but time dependent dilution flux (Eigen and Schuster, 1979)

$$\Phi(t) = \sum_{i=1}^{n} \sum_{j=1}^{n} a_{ij}\, x_i(t)\, x_j(t)$$

has been introduced for keeping the sum of relative concentrations normalized to unity at every instant. This constraint results in an internally controlled constant population size and is tantamount to a common assumption in population genetics. Since the rate constants $a_{ij}$ are non-negative the flux $\Phi(t)$ is non-negative as well. This particular form of the flux has been found to be most convenient for mathematical analysis. In addition, the qualitative dynamics is essentially the same for a continuously stirred tank reactor (CSTR), where the dilution flux is constant (Schuster and Sigmund, 1985).

By definition relative concentrations are non-negative quantities and hence equation (3) is physically meaningful on the simplex

$$S_n \doteq \left\{ (x_1, \ldots, x_n) \mid x_i \geq 0, \sum_{i=1}^{n} x_i = 1 \right\}$$

only. If $a_{ij} \geq 0$, for all $i, j = 1, \ldots n$ then $S_n$ is a compact, forward invariant set for the ODE (3).

If we choose the special case of error-free replication ($Q = E$, the identity matrix), the dynamical system (3) collapses to the *second order replicator equation*,

$$\dot{x}_k = x_k \left( \sum_{i=1}^{n} a_{ki}\, x_i - \sum_{i=1}^{n} \sum_{j=1}^{n} a_{ij}\, x_i x_j \right) \tag{4}$$

(for a review see Hofbauer and Sigmund, 1988) which we for convenience in vector notation,

$$\dot{\mathbf{x}} = \mathcal{R}(\mathbf{x}) , \tag{4'}$$

with $\mathbf{x} = (x_1, \ldots, x_n)$. The *replicator field* $\mathcal{R}(\mathbf{x}) = (\mathcal{R}_1, \ldots, \mathcal{R}_n)$ is defined in terms of its components

$$\mathcal{R}_k(\mathbf{x}) \doteq x_k \left( \sum_{i=1}^{n} a_{ki}\, x_i - \sum_{i=1}^{n} \sum_{j=1}^{n} a_{ij}\, x_i x_j \right) . \tag{5}$$

In order to make the replication-mutation equation more handy we introduce a *mean replication accuracy* $\bar{q}$, a *mean error rate* $\bar{p}$, and the *mean mutation rate* $\bar{\varepsilon}$ by

$$\bar{q} \doteq \frac{1}{n} \sum_{i=1}^{n} q_{ii}, \quad \bar{p} = 1 - \bar{q} \quad \text{and}$$

$$\bar{\varepsilon} \doteq \frac{1}{n(n-1)} \sum_{i=1}^{n} \sum_{j \neq i}^{n} q_{ij} = \frac{1}{n(n-1)} \Big( \sum_{i=1}^{n} \sum_{j=1}^{n} q_{ij} - n\bar{q} \Big) \ .$$

According to the conservation law (2), $\bar{p}$, $\bar{q}$, and $\bar{\varepsilon}$ can be converted into one another by the following expression:

$$\bar{p} = 1 - \bar{q} = (n-1)\bar{\varepsilon} \ .$$

The forthcoming analysis is facilitated by the definition of a modified mutation matrix $W \doteq (w_{ij})$ whose off-diagonal elements are weighted against the mean mutation rate, and whose diagonal elements are set to zero:

$$w_{ij} = \begin{cases} 0 & \text{for } i = j \ , \\ q_{ij} / \bar{\varepsilon} & \text{for } i \neq j \ . \end{cases}$$

The matrix $W$ is used to define a *mutation field* $\mathcal{M}(\mathbf{x}) = (\mathcal{M}_1, \ldots, \mathcal{M}_n)$ with

$$\mathcal{M}_k(\mathbf{x}) \doteq \sum_{i=1}^{n} \sum_{j=1}^{n} \big( w_{ki} \, a_{ij} \, x_i - w_{ik} \, a_{kj} \, x_k \big) \, x_j \ . \tag{6}$$

We note that the terms with $i = k$ cancel and hence the assumption of vanishing diagonal elements, $w_{kk} = 0$, has no influence on the mutation field.

The model equation (3) takes now the form

$$\dot{\mathbf{x}} = \mathcal{L}(\mathbf{x}) = \mathcal{R}(\mathbf{x}) + \bar{\varepsilon} \mathcal{M}(\mathbf{x}) \ . \tag{3'}$$

We are dealing with circulant reaction matrices $A \doteq \{a_{ij} = a_{i+1,j+1}; \, i,j \bmod n\}$ predominantly, and hence it is convenient to measure all reaction rates relative to the diagonal term which represents the autocatalytic rate constant $a_{ii} = \alpha$. This means that we split

$$\hat{a}_{ij} = \hat{a}_{ij} - \alpha, \quad \text{with } \hat{a}_{ii} = 0 \ .$$

After dropping the hats, $\hat{a}_{ij} \Rightarrow a_{ij}$, we obtain from equation (3')

$$\dot{\mathbf{x}} = \mathcal{R}(\mathbf{x}) + \bar{\varepsilon}\,\mathcal{M}(\mathbf{x}) + \bar{\varepsilon}\,\alpha\,\mathcal{B}(\mathbf{x}) \;, \tag{7}$$

where the *background field* $\mathcal{B}(\mathbf{x})$ is defined by

$$\mathcal{B}_k(\mathbf{x}) \;\doteq\; \sum_{i=1}^{n}(w_{ki}x_i - w_{ik}x_k)\;. \tag{8}$$

**Remark.** Equation (7) reduces to $\dot{\mathbf{x}} = \bar{\varepsilon}\alpha\mathcal{B}(\mathbf{x})$ in the absence of selection, i.e., if $A = \mathbf{0}$. This *mutation equation* has a unique and globally stable interior equilibrium if $\alpha\bar{\varepsilon} > 0$ and the mutation matrix $Q$ is irreducible (Hofbauer and Sigmund 1988, p.251).

**Remark.** If the mutation matrix $Q$ is bistochastic then we can use the following shorter form for mutation and background fields:

$$\mathcal{M}_k(\mathbf{x}) = \sum_{i=1}^{n} w_{ki} \sum_{j=1}^{n} x_j \left(a_{ij}x_i - a_{kj}x_k\right) \quad \text{and} \quad \mathcal{B}_k(\mathbf{x}) = \sum_{i=1}^{n} w_{ki}(x_i - x_k)\;.$$

This includes the important special cases of both *symmetric* and *circulant* mutation frequencies.

### 2.2. Existence of a central equilibrium

Since $S_n$ is compact and forward invariant for the replication-mutation equation, Brower's fixed point theorem ensures that there is always a rest point in the interior of the simplex.

　　　Many replicator equations (4) sustain an equilibrium in the interior of the simplex $S_n$. This fixed point can be readily shifted into the center $\mathbf{c} = \frac{1}{n}(1,\ldots,1)$ by a *barycentric transformation* (Hofbauer and Sigmund, 1988). It is useful to derive the conditions which leave the central fixed point unchanged.

**Lemma 2.1.** Let $\mathbf{c} = \frac{1}{n}(1,\ldots,1)$ be an equilibrium point of the replicator equation (4). Then $\mathbf{c}$ is also a fixed point of the replication-mutation equation (3) if the condition

$$WA\mathbf{1} = \frac{1}{n}(\mathbf{1}A\mathbf{1}) \cdot W\mathbf{1} \qquad (9)$$

is fulfilled.

**Proof.** Evaluation of equation (5) at the central fixed point, $\mathcal{R}(\mathbf{c}) = 0$, yields

$$n \sum_{i=1}^{n} a_{ki} = \sum_{i=1}^{n}\sum_{j=1}^{n} a_{ij},$$

or in matrix form $n(A\mathbf{1})_k = (\mathbf{1}A\mathbf{1})$ for all $k$. Substituting this into the replication-mutation equation (3) at $\mathbf{c}$,

$$\dot{x}_k(\mathbf{c}) = \frac{\bar{\varepsilon}}{n^2}\Big(\sum_{i=1}^{n}\sum_{j=1}^{n} w_{ki} a_{ij} - \sum_{\ell=1}^{n} w_{k\ell} \sum_{j=1}^{n} a_{kj}\Big),$$

and rewriting in matrix form completes the proof. ∎

**Theorem 2.1.** Suppose $A$ and $W$ are circulant. Then $\mathbf{c}$ is is a rest point of the selection mutation equation (3).

**Proof.** Let $C^l = (\delta_{i,j-l})$. The product of two such matrices fulfills $C^l C^m = C^{m+l}$ since

$$[C^l C^m]_{ki} = \sum_j \delta_{k,j-l}\delta_{j,i-m} = \delta_{k,i-m-l} = C^{l+m}_{ki}$$

where indices are defined modulo $n$. If $A$ and $W$ are circulant they can be represented by

$$A = \sum_{j=0}^{n-1} a_j C^j \qquad \text{and} \qquad W = \sum_{j=0}^{n-1} w_j C^j.$$

Then $WA = \sum_m \sum_i w_i a_{m-i} C^m$ and $C^l \mathbf{1} = \mathbf{1}$ implies

$$(WA\mathbf{1})_k = \sum_m \sum_i w_i a_{m-i} = \sum_i w_i \sum_m a_{m-i} = \sum_i w_i \sum_j a_j = (W\mathbf{1})_k \cdot (A\mathbf{1})_k$$

for arbitrary $k$. Since $A$ is circulant we have $(A\mathbf{1})_1 = \frac{1}{n}(\mathbf{1}A\mathbf{1})$, and the condition in lemma 2.1 is fulfilled. ∎

In order to create a fixed point at $\mathbf{c}$ it is by no means necessary to fulfil the condition of theorem 2. For instance, if the reaction matrix $A$ is a multiple of the identity matrix $E$, then $\mathbf{c}$ is a rest point of equation (3) for arbitrary mutation matrices $Q$.

In the forthcoming analysis we shall use different assumptions for the mutation matrix $Q$ in order to be able to derive analytical expressions for the locations and the stability properties of rest points. These assumptions are essentially related to two basic models: uniform mutation rates and uniform error rates.

### 2.3. The uniform mutation rate model

The simplest and most restrictive, as well as most unrealistic, model for mutation assumes that all mutation rates are equal, $w_{ij} = 1$ for $i \neq j$. We will refer to this mutation matrix as $\mathbf{Q}^{(1)}$:

$$\mathbf{Q}^{(1)} \doteq \left[ q_{ij} = \begin{cases} 1 - (n-1)w & \text{for } i = j \\ w & \text{for } i \neq j \end{cases} \right]$$

The corresponding $W$ matrix is the of the form $\mathbf{W}^{(1)} = \mathbf{I} - E$. Here $\mathbf{I}$ denotes the matrix with all entries equal to 1.

**Lemma 2.2.** The central point $\mathbf{c}$ is an equilibrium of the replication-mutation equation (7) with mutation matrix $\mathbf{Q}^{(1)}$ if $\mathbf{c}$ is an equilibrium of the replicator equation (4).

**Proof.** Substituting $\mathbf{Q}^{(1)} = \mathbf{I} - E$ into lemma 2.1 yields

$$\mathbf{I}A\mathbf{1} - A\mathbf{1} = (\mathbf{1}A\mathbf{1})\mathbf{1} - A\mathbf{1} = (\mathbf{1}A\mathbf{1})\mathbf{1} - \frac{1}{n}(\mathbf{1}A\mathbf{1})\mathbf{1},$$

i.e., we need $(A\mathbf{1})_k = \frac{1}{n}(\mathbf{1}A\mathbf{1})$ for all $k$. This condition is necessary and sufficient for $\mathbf{c}$ to be an equilibrium of the replicator equation (4). ∎

*2.4. The uniform error rate model*

The model assume that a species $\mathbf{I}_k$ is represented by a sequence of length $\nu$. We further assume that mutants appear through copying one or more symbols incorrectly, and that the error rates are independent of the position of the digit in the sequence (Eigen, 1971; Eigen and Schuster, 1977; for a recent survey of the uniform error rate model see Eigen *et al.*, 1988 and 1989). The fidelity $q$ of the replication process is a measure of the probability that a single symbol is copied correctly. The *Hamming distance* $d(i,k)$ of two sequences $\mathbf{I}_i$ and $\mathbf{I}_k$ counts the number of digits in which they differ (Hamming, 1950). Binary sequences of length $\nu$ can be readily mapped on a hypercube such that the Hamming distance of the sequences counts the minimum number of edges that have to be passed in order to go from one sequence to the other. For short we call such a model the hypercube model and we obtain a specific form for the mutation matrix $Q$, which we shall denote by $\mathbf{Q}^{(2)}$:

$$\mathbf{Q}^{(2)} \doteq \left\{ q_{ij} = q^\nu \left(\frac{1-q}{q}\right)^{d(i,j)} \right\}.$$

For convenience we introduce $b \doteq \frac{1-q}{q}$ and the matrix $B(\nu)$: $b_{ij} \doteq b^{d(i,j)}$ for a $\nu$-digit sequence. $B(\nu)$ is a $2^\nu \times 2^\nu$ matrix. The B-matrices can be obtained by recursion (Rumschitzky, 1987):

$$B(1) \doteq \begin{pmatrix} 1 & b \\ b & 1 \end{pmatrix} \qquad B(\nu+1) \doteq \begin{pmatrix} B(\nu) & bB(\nu) \\ bB(\nu) & B(\nu) \end{pmatrix}$$

The eigenvalues of $B(1)$ are $\lambda_\pm = 1 \pm b$ with the eigenvectors $\mathbf{x}_\pm = (1, \pm 1)$. The eigenvectors $\mathbf{x}_-$ and $\mathbf{x}_+$ are orthogonal. Let us assume that $\mathbf{x}_1$ is an eigenvector of $B(\nu)$ with the eigenvalue $\lambda(\nu)$. Then $\mathbf{y}_+ = (\mathbf{x}_1 \oplus \mathbf{x}_1)$ and $\mathbf{y}_- = (\mathbf{x}_1 \ominus \mathbf{x}_1)$ are eigenvectors of $B(\nu+1)$ and for the eigenvalues holds $\lambda(\nu+1)_\pm = (1 \pm b)\lambda(\nu)$. Thus we can write down the eigenvalues of $B(\nu)$ explicitly:

$$\lambda_j(\nu) = (1+b)^{\nu-j}(1-b)^j, \qquad j = 0, \ldots, \nu$$

where $\lambda_j(\nu)$ is $\binom{\nu}{j}$-fold degenerate. The above procedure produces a complete set of orthogonal eigenvectors of $B(\nu)$. In particular, if we choose the +-sign in each recursion step we find that **1** is a simple eigenvector of $B(\nu)$ for all dimensions. Of course $B(\nu)$ and $\mathbf{Q}^{(2)}(\nu)$ have the same eigenvectors. The eigenvalues of $\mathbf{Q}^{(2)}(\nu)$ are readily obtained:

$$\phi_j = (2q-1)^j. \tag{10}$$

### 2.5. The single point mutation model

The hypercube model is simplified further by the assumption that error rates are so small that only single point mutations occur frequently enough to be considered. This is tantamount to assuming ($\overline{p} = 1 - \overline{q} = (n-1)\overline{\varepsilon}$):

$$\overline{q}^2 = (1-\overline{p})^2 \approx 1 - 2\overline{p} \to 1 - 2w$$

$$\overline{q}(1-\overline{q}) = (1-\overline{p})\overline{p} \approx \overline{p} \to w$$

$$(1-\overline{q})^2 = \overline{p}^2 \approx 0 \to 0$$

Mutations are thus restricted to the interconversion of species with Hamming distance $d(i,j) = 1$. The recursion reads:

$$\tilde{B}(\nu+1) \doteq \begin{pmatrix} \tilde{B}(\nu) & b \cdot E \\ b \cdot E & \tilde{B}(\nu) \end{pmatrix}, \qquad \tilde{B}(1) \doteq B(1)$$

We find, that the eigenvectors are the same for $B(\nu)$ and $\tilde{B}(\nu)$. For the eigenvalues we find

$$\tilde{\lambda}_j(\nu) = 1 + (\nu - 2j)b, \qquad j = 0, \ldots, \nu$$

Again the eigenvalue $\lambda_j$ is $\binom{\nu}{j}$-fold degenerate. For the matrix $\mathbf{Q}^{(3)\prime} \doteq q^\nu \cdot B$ we have $\sum_{j=1}^{2^\nu} = q^\nu(1+\nu b) = q^\nu + \nu q^{\nu-1}(1-q)$. In order to keep the sum of mutation probabilities normed to one, we define:

$$\mathbf{Q}^{(3)} \doteq \frac{1}{q^\nu + \nu q^{\nu-1}(1-q)} \cdot \mathbf{Q}^{(3)\prime}$$

We can now easily calculate also the eigenvalues $\tilde{\phi}_j$ of $\mathbf{Q}^{(3)}(\nu)$.

$$\tilde{\phi}_j = \frac{q + (\nu - 2j)(1-q)}{q + \nu(1-q)}. \tag{11}$$

### 3. Models with two species

*3.1. The general case*

The (bistochastic) mutation matrix of a replication-mutation-system with two species is uniquely determined,

$$Q = \begin{pmatrix} 1 - \overline{\varepsilon} & \overline{\varepsilon} \\ \overline{\varepsilon} & 1 - \overline{\varepsilon} \end{pmatrix}, \quad \text{with} \quad 0 \leq \overline{\varepsilon} \leq 1.$$

The most general reaction matrix is

$$A = \begin{pmatrix} a & b \\ c & d \end{pmatrix} \quad \text{with} \quad a, b, c, d > 0.$$

The selection mutation equation is defined on the unit interval. We can eliminate the concentration of species $\mathbf{I}_2$, $x_2$, by setting $x \doteq x_1 = 1 - x_2$. We obtain

$$\begin{aligned}\dot{x} =& [-a + b + c - d]x^3 + [a - 2b - c + 2d + (-a + b - c + d)\overline{\varepsilon}]x^2 + \\ & + [b - d + (-b + c - 2d)\overline{\varepsilon}]x + d\overline{\varepsilon} \qquad \doteq f(x)\end{aligned} \quad (12)$$

The rest points of equation (12) can be obtained by solving the cubic equation $f(x) = 0$. It turns out, however, that the resulting expressions are too complicated to be useful for further analytical work.

The mutation free system has been analyzed completely by (Eigen & Schuster, 1978). There are four generic phase portraits

$a > c \wedge d > b$     Competition of $\mathbf{I}_1$ and $\mathbf{I}_2$ (bistability).

$a > c \wedge d < b$     Selection of $\mathbf{I}_1$, i.e., one equilibrium at $x_1 = 1$.

$a < c \wedge d > b$     Selection of $\mathbf{I}_2$, i.e., one equilibrium at $x_2 = 1$.

$a < c \wedge d < b$     Cooperation of $\mathbf{I}_1$ and $\mathbf{I}_2$.

The phase diagram for the mutation free case is show in figure 1 (a) in terms of the parameters

$$r = -a + b + c - d \qquad \text{and} \qquad s = a + b - c - d$$

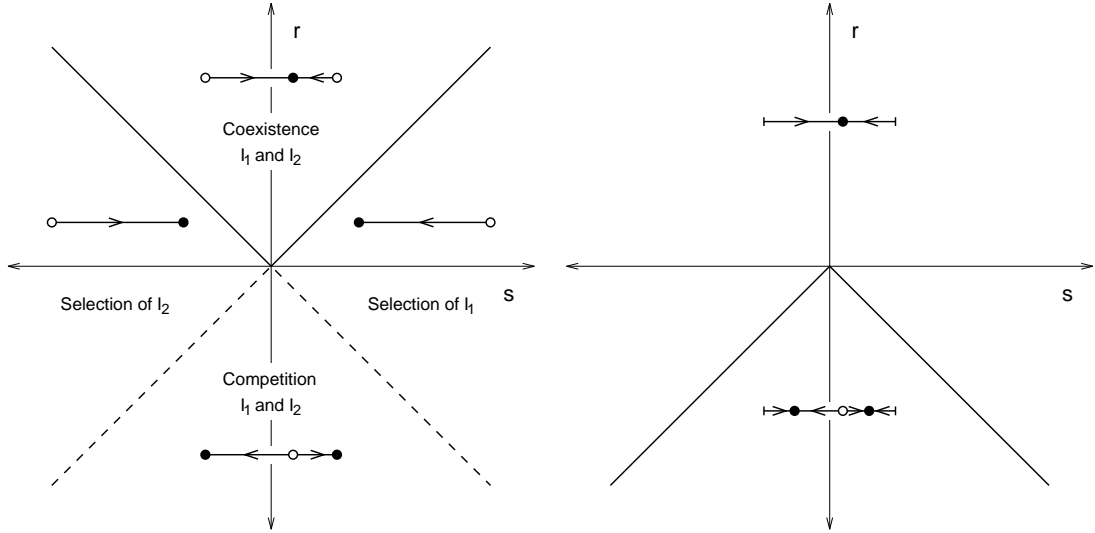

**Figure 1:** Phase diagram of the two species model. **(a)** for $\bar\varepsilon=0$, **(b)** for small $\bar\varepsilon>0$.

It is possible to generalize these results to small mutation rates using a perturbation approach (Stadler and Schuster, 1992), see figure 1 (b).

If $a = d = 0$ then $x = 0$ and $x = 1$ are rest points of equation (3') for all $\bar\varepsilon$. It can be shown analogous to the proof of lemma 3.2 below that they are always unstable. In the following we will assume that $(a, d) \neq (0, 0)$. We will use the following abbreviations for the coefficients of $f(x)$:

$$\begin{aligned} a_3 &= -a + b + c - d \quad = r \\ a_2 &= a - 2b - c + 2d + [-a + b - c + d]\bar\varepsilon \\ a_1 &= b - d + [-b + c - 2d]\bar\varepsilon \\ a_0 &= d\bar\varepsilon \end{aligned} \qquad (13)$$

**Lemma 3.1.** Let $\bar\varepsilon > 0$. Then there are only two possible generic phase portraits consisting of

[A] a unique globally stable rest point, or

[B] two locally stable rest points separated by an unstable one

The bifurcation between the two phase portraits is generically a saddle-node bifurcation.

**Proof.** The unit interval is strictly forward invariant, hence there is at least one stable rest point. If it is degenerate, i.e. in the case it has a zero eigenvalue, then we have a pitchfork bifurcation. Since $f(x)$ is cubic, there are at most three equilibria and at most one of them can be degenerate. If a rest point is degenerate, then there are at most two equilibria (this situation corresponds to a saddle-node bifurcation, or in a degenerate case to a transcritical bifurcation). If there are exactly two equilibria in $(0,1)$ then continuity ensures that the flow changes its direction at only one of them, hence the second rest point is degenerate. This situation is of course not generic since an arbitrarily small perturbation of any of the coefficients $a_i$ leads to a phase portrait with either one or three rest points. If we have three rest points in $(0,1)$ then none of them can be degenerate and, by continuity, two of them are stable and the unstable equilibrium lies between them.

∎

**Lemma 3.2.** If $r$ is non-negative, then equation (12) has phase portrait **[A]** for all $\overline{\varepsilon} > 0$.

**Proof.** If $a_3 > 0$, then equation (12) has either a single unstable rest point somewhere on the real line ($f' \geq 0$ everywhere), or it has three rest points on the real line, two which are unstable ($f$ has then a local maximum and a local minimum. Lemma 3.1 guarantees that the first possibility can never occur for a selection mutation equation. For the second case we know (again from lemma 3.1) that there must be a single stable equilibrium in $(0,1)$. A degenerate rest point occurs if $f(x) = 0$ for the local minimum or the local maximum of $f$. In both cases the second rest point must be unstable, contradicting lemma 3.1. For the special case $a_3 = 0$ we have at most two rest points on the real line. Suppose $f(x) = 0$ has two real solutions, then one of them corresponds to a stable rest point, the other one to an unstable one. Lemma 3.1 guarantees that we have phase portrait [**A**]. More degenerate cases cannot occur since we have always a stable rest point in $(0,1)$. ∎

**Lemma 3.3.** A necessary condition for the occurrence of phase portrait [**B**] is

$$
\begin{aligned}
&\text{(i)} \quad -3a_3 > a_2 > 0, \text{ and} \\
&\text{(ii)} \quad a_1 + a_2 > 0.
\end{aligned}
\tag{14}
$$

**Proof.** For two stable equilibria existing in the unit interval it is necessary that $f'(x)$ has its maximum $x_m$ in $(0,1)$ and that $f'(x_m) > 0$, since there must be an unstable equilibrium in $(0,1)$. From

$$f''(x_m) = 6a_3 x_m + 2a_2 = 0$$

we find $x_m = -a_2/3a_3$. Lemma 3.2 implies that $a_3 < 0$ is fulfilled. Then $x_m \in (0,1)$ implies condition (i). A short calculation shows that $f'(x_m) = a_2 x_m + x_1$. Since $0 < a_2 x_m < a_2$, it is necessary that at least $a_2 + a_1 > 0$. ∎

**Corollary.** There is an unique rest point for

$$\overline{\varepsilon} \geq 1 - \frac{b+c}{a+d} \tag{15}$$

**Proof.** A sufficient condition for ruling out phase portrait [**B**] is $a_1 + a_2 = (a+d) - (b+c) - \overline{\varepsilon}(a+d) < 0$. Equation (15) is an immediate consequence. ∎

Hence there is a critical mutation rate $\overline{\varepsilon}^*(a,b,c,d) < 1$ such that there is a unique and globally stable equilibrium for all $\overline{\varepsilon} > \overline{\varepsilon}^*$. Equation (15) provides an upper bound on $\overline{\varepsilon}^*$. This bound is reasonable, in the sense that it reproduces the condition $a_3 < 0$ which is necessary for the existence of phase portrait [**B**], but which is far from being optimal.

**Remark.** For $\bar{\varepsilon} = 1/2$ we find that $x = 1/2$ is a rest point for all coefficients $a, b, c, d$.

The bifurcation points of equation (12) as a function of $\bar{\varepsilon}$ can be found in principle by solving

$$f(x) = 0 \qquad \wedge \qquad f'(x) = 0.$$

In case of the general model we obtain a quartic equation for $\bar{\varepsilon}$ with no simple solutions.

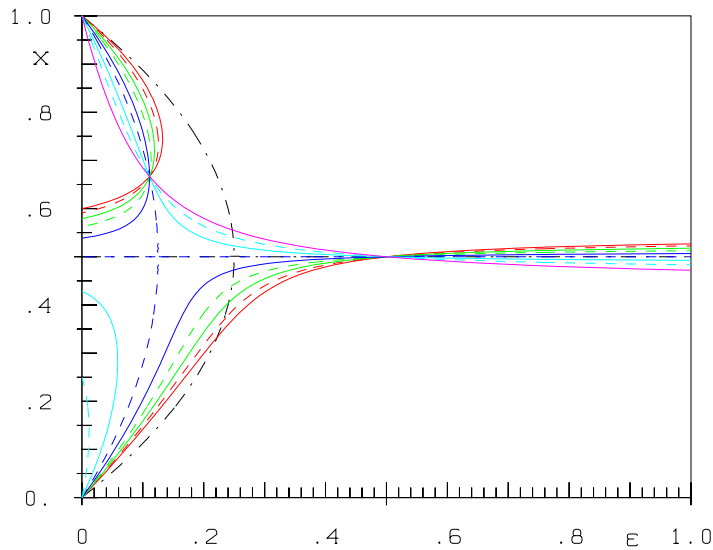

**Figure 2:** Numerical bifurcation diagram showing the solutions of $f(x)=0$ for $a=10, b=5$, $d=2(10-c)$, and $0 \leq c \leq 10$ as a function of $\bar{\varepsilon}$. Each choice of $c$ is shown in its line style. The dashed parabola belongs to the symmetric case $c=5$, for which we observe a pitchfork bifurcation. The dot-dashed parabola $\bar{\varepsilon}=1/4-(1/2-x)^2$ defines the boundary of the region within which we found more than one rest point in all our numerical examples.

For small mutation rates we need $a > c$ and $d > b$ for the existence of three equilibria. Numerical solutions of $f(x) = 0$ indicate that these conditions are necessary for arbitrary $\bar{\varepsilon}$. We have calculated the bifurcation diagrams of such systems as a function of $\bar{\varepsilon}$ for a large number of choices for $a, b, c, d$ (see figure 2 for an example). Our numerical experiments suggest the following

**Conjecture.** $\bar{\varepsilon}^*(a,b,c,d) \leq 1/4$.

*3.2. Special cases*

The additional constraint $a+b = c+d$ ensures that **c** is a rest point, i.e, $f(1/2) = 0$, for all $\bar{\varepsilon}$. A rather lengthy calculation shows that this rest point is stable if

$$\bar{\varepsilon} > \frac{1}{4}\left(1 - \frac{b+c}{a+d}\right)$$

Even for this simplified model we did not succeed in finding a better bound for $\bar{\varepsilon}^*$.

The special case of a symmetric reaction matrix

$$A = \begin{pmatrix} a & b \\ b & a \end{pmatrix} \qquad (16)$$

contains both the hypercycle $A = C^1$ and the two-dimensional Schlögl model $A = E$ as special cases. It can be analyzed in full detail.

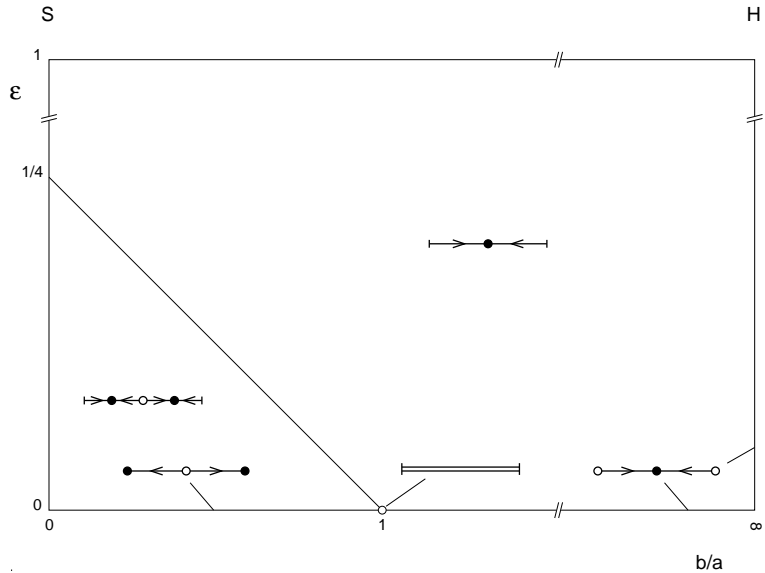

**Figure 3:** The parameter space for the symmetric two species model (16). Shown are different phase-portraits giving rise to qualitatively different dynamics for the parameters $\frac{b}{a}$ and $\bar{\varepsilon}$ (Note: at $\frac{b}{a}=1$ and $\bar{\varepsilon}=0$ every point in [0,1] is a rest point). $S$ and $H$ denote the Schlögl model and the hypercycle model, respectively.

**Table 1.** Eigenvalues of the Jacobian at the fixed points.

|             | $\hat{x}_1$               | $\hat{x}_2, \hat{x}_3$      |
|-------------|---------------------------|------------------------------|
| $\lambda_0$ | $-\frac{1}{2}$            | $a(2\overline{\varepsilon} - 1)$ |
| $\lambda_1$ | $\frac{a-b}{2} - 2a\overline{\varepsilon}$ | $(b - a) + 4a\overline{\varepsilon}$ |

**Lemma 3.4.** The symmetric model, equation (16), has an unique and globally stable equilibrium, i.e, phase portrait **[A]** if

$$\overline{\varepsilon} \geq \overline{\varepsilon}^* = \frac{1}{4}\left(1 - \frac{b}{a}\right).$$

For $\overline{\varepsilon} < \overline{\varepsilon}^*$ we have phase portrait **[B]**.

**Proof.** Equation (12) reduces to

$$\dot{x} = 2(b - a)x^3 - 3(b - a)x^2 + (b - 2a\overline{\varepsilon} - a)x + a\overline{\varepsilon}. \tag{17}$$

with the equilibria

$$x_0 = \frac{1}{2}, \qquad x_\pm = \frac{1}{2} \pm \sqrt{\frac{1}{4} - \frac{a\overline{\varepsilon}}{a - b}} \tag{18}$$

The square root in equation (18) is real and non-zero for $\overline{\varepsilon} < (1 - b/a)/4$, and and $x_\pm$ lies in $[0, 1]$, hence we have three fixed points. For $\overline{\varepsilon} = \overline{\varepsilon}^*$ the square root vanishes and $x_\pm = x_0 = 1/2$, i.e., there is only one rest point which must be stable by lemma 1. For $\overline{\varepsilon} > \overline{\varepsilon}^*$ the equilibrium $x = 1/2$ is unique. ∎

Lemma 3.4 shows that that we cannot find a better uniform bound for $\overline{\varepsilon}^*$ than the conjecture $\overline{\varepsilon}^* = 1/4$.

It is straightforward to compute explicitly the Jacobian and its eigenvalues at the positions of the fixed points (Table 1). The eigenvalue $\lambda_0$ belongs to the external eigenvector which does not influence the dynamical behaviour.

With $\bar{\varepsilon} = \bar{\varepsilon}^*$ and $y = x - 1/2$ we find that equation (17) reduces to

$$\dot{y} = \frac{1}{4}(b-a)y^3$$

The analytical solution of this ODE reads:

$$y(t) = \text{sgn}(y(0)) \cdot \frac{1}{\sqrt{\frac{1}{y(0)^2} + \frac{1}{2}(a-b)t}}$$

and $(\frac{1}{2}, \frac{1}{2})$ is stable if $a \neq b$. If $a = b$ we have $\bar{\varepsilon}^* = 0$, i.e., we recover the mutation free replicator equation with all entries in the reaction matrix being equal. In this case all points are rest points.

Figure 3 shows the regimes of phase portraits in the two-dimensional space of the parameters $\bar{\varepsilon}$ and $\frac{b}{a}$. For $\frac{b}{a} \to 0$ we have the Schlögl model and for $\frac{b}{a} \to \infty$ we have the hypercycle as limiting case. Obviously this parametrization is sufficient to describe our model. The parameter $\frac{b}{a}$ measures the cooperativity in the system: the Schlögl model refers to no cooperativity ($\frac{b}{a} = 0$), and the hypercycle is "cooperativity pure" ($\lim \frac{b}{a} \to \infty$).

### 4. The hypercycle model

The reaction matrix $A$ of the elementary hypercycle (Eigen and Schuster, 1978) is given by $C^1$, i.e., by

$$a_{ij} = \delta_{i,j-1}; \quad i,j = 1, 2, \ldots, \text{mod} \, n,$$

and represents a special case of a (non-symmetric) circulant matrix. The corresponding replicator equation has an asymptotically stable rest point in the interior of the simplex $S_n$ for $n = 2, 3$, and $4$. For $n \geq 5$ the central fixed point becomes unstable and there is an asymptotically stable limit cycle in the interior of $S_n$ (Hofbauer *et al.*, 1991).

The replication field $\mathcal{R}(\mathbf{x})$ and the mutation field $\mathcal{M}(\mathbf{x})$ in the hypercycle model simplify to

$$\begin{aligned}\mathcal{R}_k(\mathbf{x}) &= x_k\left(x_{k-1} - \sum_{\ell=1}^n x_\ell x_{\ell-1}\right) \quad \text{and} \\ \mathcal{M}_k(\mathbf{x}) &= \sum_{\ell=1}^n w_{k\ell}\left(x_\ell x_{\ell-1} - x_k x_{k-1}\right),\end{aligned} \quad (19)$$

where all indices are understood mod $n$. The background field $B(\mathbf{x})$ remains unchanged. By theorem 2.1, $\mathbf{c}$ is an equilibrium of the replication-mutation equation of elementary hypercycles for arbitrary $n$, $\alpha$, $\overline{\varepsilon}$ and $W$.

As shown in (Stadler and Schuster, 1992) there are no other rest points for small $\overline{\varepsilon} > 0$. Numerical studies indicate that the hypercycle model with mutation has in fact a unique globally stable equilibrium for all $\overline{\varepsilon} > 0$. In the following we will restrict our attention the the rest point $\mathbf{c}$.

The Jacobian at the central equilibrium, $D \doteq \left(d_{ij} = \partial \mathcal{L}_i/\partial x_j(\mathbf{c})\right)$, is readily computed. Its elements $d_{k\ell}$ are of the form

$$d_{k\ell}(\mathbf{c}) = \frac{1}{n}\left(\delta_{k-1,\ell} - \frac{2}{n}\right) + \frac{\overline{\varepsilon}}{n}\Big(w_{k\ell} + w_{k,\ell+1} - (n-1)(\delta_{k,\ell} + \delta_{k-1,\ell})\Big) + \\ + \alpha\overline{\varepsilon}\Big(w_{k\ell} - (n-1)\delta_{k,\ell}\Big). \quad (20)$$

The properties of the central fixed point of elementary hypercycles with mutation are summarized in several lemmas and theorems.

**Lemma 4.1.** $D$ is circulant if and only if $W$ is circulant.

**Proof.** Define $Y_{kl} \doteq (\alpha + \frac{1}{n})w_{kl} + \frac{1}{n}w_{k,l+1}$ and

$$Z_{kl} \doteq \frac{1}{n}(\delta_{k-1,l} - \frac{2}{n}) + \frac{1-n}{n}\overline{\varepsilon}(\delta_{k,l} + \delta_{k-1,l}) - \alpha\overline{\varepsilon}(n-1)\delta_{k,l}.$$

We have $D = Z + \overline{\varepsilon}Y$. $Z$ is circulant since it is a linear combination of the matrices $E = (\delta_{ij})$ and $C^1 = (\delta_{i-1,j})$ which are both circulant. Clearly, $D$ is circulant if and only if $Y$ is circulant. It is obvious that $Y$ is circulant if $W$ is circulant. It remains to show the converse, i.e., suppose $Y$ is circulant. Then

$$Y_{1l} = (\alpha + \frac{1}{n})w_{1,l} + \frac{1}{n}w_{1,l+1} = (\alpha + \frac{1}{n})w_{1+j,l+j} + \frac{1}{n}w_{1+j,l+1+j} = Y_{1+j,l+j}$$

must hold. Setting $l = 1$ and using $w_{11} = 0$ we find $w_{12} = w_{1+j,2+j}$ for all $j$. Now we proceed by induction; suppose $w_{1,l} = w_{1+j,l+j}$. $W$ is circulant if and only if

$$(\alpha + \frac{1}{n})w_{1,l} - (\alpha + \frac{1}{n})w_{1+j,l+j} = \frac{1}{n}w_{1,l+1} - \frac{1}{n}w_{1+j,l+1+j} = 0.$$

This simplifies to $w_{1,l+1} = w_{1+j,l+1+j}$, i.e., $W$ must be circulant. ∎

**Lemma 4.2.** If $W$ is circulant then the real parts $\gamma_m$ of the eigenvalues $\phi_m$ of $D$ for $m = 1, \ldots, n-1$ are given by

$$\begin{aligned}\gamma_m &= \frac{1}{n}\cos\frac{2\pi m}{n} - \overline{\varepsilon}\frac{n-1}{n}(1 + \cos\frac{2\pi m}{n}) - \alpha\overline{\varepsilon}(n-1) \\ &+ \frac{\overline{\varepsilon}}{n}\sum_{i=1}^{n}(w_{1,j} + w_{1,j+1})\cos\frac{2\pi m(j-1)}{n} + \overline{\varepsilon}\alpha\sum_{j=2}^{n}w_{ij}\cos\frac{2\pi m(j-1)}{n}\end{aligned} \quad (21)$$

The remaining eigenvalue $\phi_0 = -1/n$ belongs to the eigenvector $\mathbf{1}$ and does not influence the dynamics.

**Proof.** Let $c_j \doteq d_{1+k,j+k}$. If $D$ is an arbitrary circulant matrix, the eigenvalues of $D$ are given by $\phi_m = \sum_{j=1}^{n} c_j \lambda_n^{m(j-1)}$ where $\lambda_n \doteq \exp(\frac{2\pi}{n}i)$. Hence

$$\gamma_m = \Re\phi_m = \sum_{i=1}^{n} c_j \Re\lambda_n^{m(j-1)} = \sum_{i=1}^{n} c_j \cos\frac{2\pi m(j-1)}{n}$$

A rather long but straight forward calculation then yields equation (21). In order to check that $\mathbf{1}$ is an eigenvector we calculate

$$[D \cdot \mathbf{1}]_k = \sum_{l=1}^{n}\left[\frac{1}{n}(\delta_{k-1,l} - \frac{2}{n}) + \frac{\overline{\varepsilon}}{n}(w_{kl} + w_{k,l+1} - (n-1)(\delta_{k,l} + \delta_{k-1,l})) + \right.$$
$$\left. + \alpha\overline{\varepsilon}(w_{kl} - (n-1)\delta_{k,l})\right] =$$
$$= -\frac{1}{n} = \gamma_0$$

This eigenvector **1** points outside the simplex. Since $S_n$ is invariant for equation (4), the dynamical behaviour of our system is determined by the remaining $n-1$ eigenvectors $\phi_1, \ldots, \phi_{n-1}$. ∎

**Theorem 4.1.** If the mutation matrix $W$ is circulant, the central fixed point of the 2-, 3- and 4-species hypercycle is stable for $\alpha, \overline{\varepsilon} \geq 0$. In the case $n = 4$, $\alpha = 0$ we need the additional condition $w_{12} \neq 3$, i.e. more then one mutation probability $q_{ij}, i \neq j$ must be nonzero.

**Proof.**

(1) $n = 2$ The dynamical behaviour is determined by the eigenvalue

$$\gamma_1 = -\frac{1}{2} - \frac{\overline{\varepsilon}}{2} w_{12} - 2\alpha\overline{\varepsilon}$$

Obviously $\gamma_1 < 0$ holds under the conditions of the theorem, therefore **c** is stable.

(2) $n = 3$ We get one double eigenvalue

$$\gamma_1 = \gamma_2 = -\frac{1}{3} + \frac{\overline{\varepsilon}}{3}(\frac{3}{2} w_{12} - 3)$$

Since $w_{12} \leq 2$ the bracket is nonpositive under our conditions and we find $\gamma_1 < 0$.

(3) $n = 4$ In this case we get one single and one double eigenvalue

$$\gamma_1 = \gamma_3 = -\frac{\overline{\varepsilon}}{2}(w_{12} - 3) - \alpha\overline{\varepsilon}(3 + w_{13})$$

$$\gamma_2 = -\frac{1}{4} - 2\alpha\overline{\varepsilon}(3 - w_{13})$$

Using $w_{ij} \leq 3$ we have $\gamma_2 \leq -\frac{1}{4}$ and with $w_{12} < 3$ $\gamma_1$ also remains negative for $\overline{\varepsilon} > 0$. For the non-mutating four membered hypercycle ($\overline{\varepsilon} = 0$) we know that the inner fixed point is stable (Eigen and Schuster, 1978; Schuster *et al.*, 1978). ∎

For larger hypercycles the expressions get much more complicated due to the increasing number of mutation terms. We therefore restrict our investigation to the special mutation matrix $\mathbf{Q}^{(1)}$.

**Theorem 4.2.** Let $w_{ij} = 1$ for $i \neq j$ and $n \geq 5$. Then bifurcations at the central fixed point **c** occur at $\overline{\varepsilon} = \overline{\varepsilon}_m(\alpha; n)$ with

$$\overline{\varepsilon}_m(\alpha; n) \doteq \frac{1}{n} \cdot \frac{\cos \frac{2\pi m}{n}}{1 + \cos \frac{2\pi m}{n} + n\alpha} \qquad 1 \leq m \leq \frac{n}{4} \tag{22}$$

The central equilibrium **c** is stable if $\overline{\varepsilon} > \overline{\varepsilon}_1(\alpha; n)$. If $n \equiv 0 \pmod{4}$ then there is a bifurcation for $\overline{\varepsilon} = 0$.

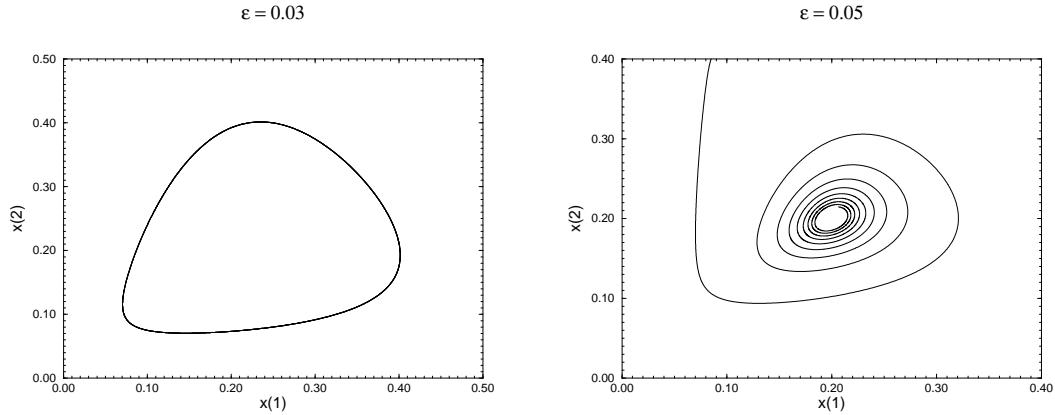

**Figure 4:** Bifurcation of the hypercycle with n=5. The limit cycles on the lhs corresponds to error rates $\overline{\varepsilon}$ below the critical value. The trajectory on the rhs represents a typical solution curve above the critical value. It spirals inwards and ultimately reaches the stable rest point in the center of $S_5$.

**Proof.** Because of the simple form of $\mathbf{Q}^{(1)}$ expression (22) simplifies considerably. Substituting the identities

$$\sum_{j=2}^{n} \cos \frac{2\pi m(j-1)}{n} = -1, \qquad \sum_{j=1}^{n-1} \cos \frac{2\pi m(j-1)}{n} = -\cos \frac{2\pi m}{n}$$

into equation (22) yields

$$\gamma_m = \frac{1}{n} \cos \frac{2\pi m}{n} - \overline{\varepsilon}[1 + \cos \frac{2\pi m}{n} + \alpha n]$$

The expression in brackets is always non-negative. Thus $\gamma_m$ is always negative if $\cos \frac{2\pi m}{n} \leq 0$. Therefore we find simple zeros of $\gamma_m$ for $1 \leq m \leq \frac{n}{4}$ at $\overline{\varepsilon} = \overline{\varepsilon}_m(\alpha;n)$. For $1 \leq m \leq n-1$ we have $\cos \frac{2\pi}{n} \geq \cos \frac{2\pi m}{n}$, therefore $\overline{\varepsilon}_1(\alpha;n) \geq \overline{\varepsilon}_m(\alpha;n)$ holds for $1 \leq m \leq \frac{n}{4}$. Since **c** is stable if all eigenvalues $\gamma_m$ are negative, we find that $\overline{\varepsilon} > \overline{\varepsilon}_1(\alpha;n)$ is sufficient condition for stability.

If $n \equiv 0 \mod 4$ then $\overline{\varepsilon}_{n/4}(\alpha;n) = 0$. ∎

We remark that $\phi_m$ and $\phi_{n-m}$ form a pair of conjugate complex eigenvalues if and only if $m \neq \frac{n}{2}$: $\phi_m = \gamma_m + i \cdot \overline{\varepsilon} \frac{n-1}{n} \sin \frac{2\pi m}{n}$ Therefore the bifurcations at $\overline{\varepsilon}_m(\alpha;n)$ with $1 \leq m < \frac{n}{4}$ are Hopf bifurcations. Numerical work for $n = 5$ shows that the bifurcation at $\overline{\varepsilon} = \overline{\varepsilon}_1(\alpha;5) \approx 0.047213$ is supercritical figure 4.

We mention further that a rigorous proof for the existence of asypmtotically stable limit cycles in the elementary hypercycle was already hard to derive for the error-free system (Hofbauer *et al.*, 1991).

### 5. The generalized Schlögl model

In Schlögl's model the replication matrix is defined to have only diagonal terms $a_{ij} \doteq a_i \cdot \delta_{ij}$. We consider here the simpler case of equal rate constants ($a_{ij} \doteq \delta_{ij}$). From Lemma 2.1 follows that **c** is a fixed point of equation (7) for arbitrary $\alpha, \overline{\varepsilon}$ and $W$. Because of the simple form of $A$, it is more convenient to use equation (7) than equation (3). We have

$$\dot{x}_k = \sum_{i=1}^{n} q_{ki} x_i^2 - x_k \sum_{j=1}^{n} x_j^2 - \alpha \sum_{i=1}^{n} q_{ki}(x_i - x_k) \qquad (23)$$

The case of two species has been treated already in section 3.2.

5.1. *Models with three species*

The calculations for the general three species case are rather sophisticated and thus we can treat here only the most simple model

$$A = \begin{pmatrix} 1 & 0 & 0 \\ 0 & 1 & 0 \\ 0 & 0 & 1 \end{pmatrix}, \qquad Q = \begin{pmatrix} 1 - 2\overline{\varepsilon} & \overline{\varepsilon} & \overline{\varepsilon} \\ \overline{\varepsilon} & 1 - 2\overline{\varepsilon} & \overline{\varepsilon} \\ \overline{\varepsilon} & \overline{\varepsilon} & 1 - 2\overline{\varepsilon} \end{pmatrix}$$

In this case the coordinates of all fixed points can be calculated for arbitrary $\overline{\varepsilon}$. From symmetry reasons, we expect three types of fixed points:

(i) a unique inner fixed point $\frac{1}{3}\mathbf{1}$

(ii) three edge-type fixed points $(x_+, x_+, y_-)$

(iii) three corner-type fixed-points $(x_-, x_-, y_+)$

The analytical results, which were obtained by using of the software package REDUCE, are compiled in table 2. We find three types of phase portraits in this system. For $0 < \overline{\varepsilon} < \frac{1}{6}$ we have a smooth deformation of phase portrait of the Schlögl model, for $\overline{\varepsilon} > 3 - 2\sqrt{2}$ there is a unique interior sink. Between $\frac{1}{6}$ and $3 - 2\sqrt{2} \approx 0.17157$ we find a rather strange phase portrait with four coexisting sinks, and three very narrow channels connecting small zones at the boundary with the central sink (figure 5).

We find that the sinks at the corners and the saddles on the edges move inwards with incerasing error rate. The saddles collide with the interior source at $\overline{\varepsilon} = \frac{1}{6}$. As $\overline{\varepsilon}$ increases further, the interior fixed point is left as sink, which is separated by the three edge-type saddles from the other three corner-type sinks. When $\overline{\varepsilon}$ reaches $3 - 2\sqrt{2}$, the saddles collide with the corner-type sinks and annihilate. If $\overline{\varepsilon}$ increases further, we are left with a unique stable fixed point. We illustrate this behaviour in a *constraint-response* plot in (figure 6),

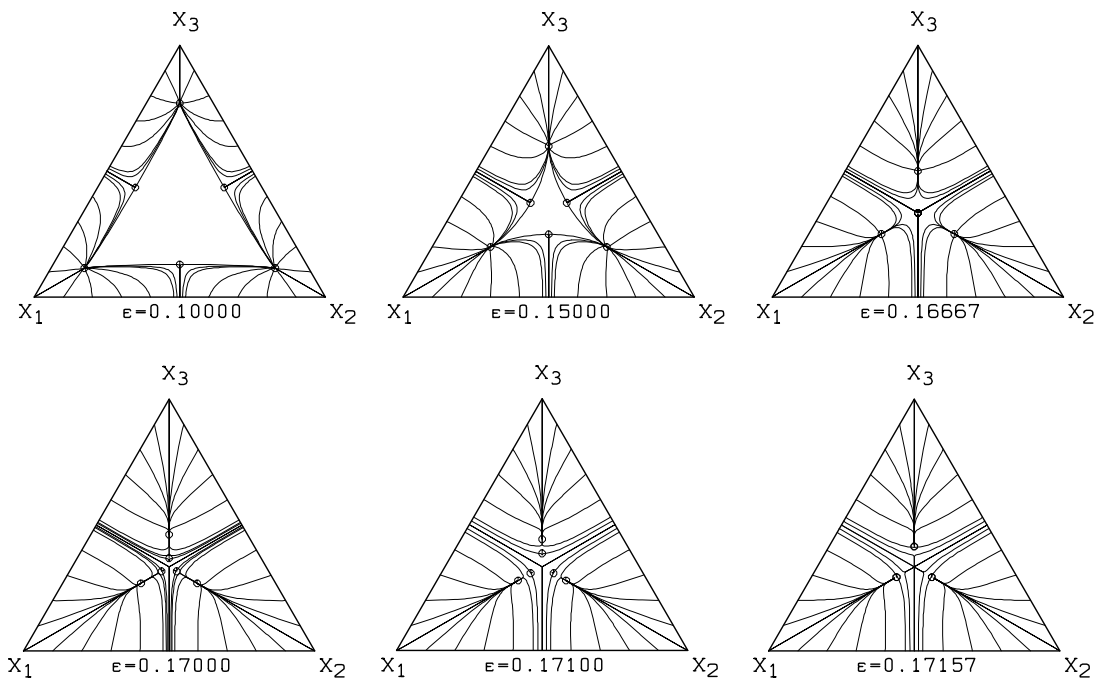

**Figure 5:** Fixed points and selected trajectories for the tree-species Schlögl model. All trajectories start at the boundary of the simplex.

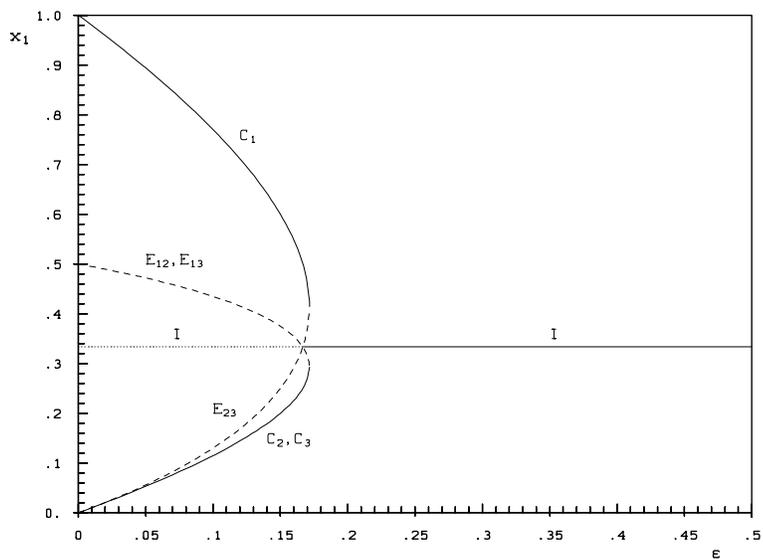

**Figure 6:** Bifurcation diagram for the rest points of the model ($I$ is the interior equilibrium, $C$ denotes rest points evolving out of the corners of the simplex, and $E$ labels fixed points starting at the edge of the triangle). Solid line denote sinks, dashed lines mark saddles, and dotted lines stand for sources.

**Table 2.** Fixed points and signs of the eigenvalues for the three-species Schlögl model

| Type | Interior | Edge | Corner |
|---|---|---|---|
| Coordinates | $\frac{1}{3}\mathbf{1}$ | $(x_+, x_+, y_-)$ | $(x_-, x_-, y_+)$ |
| | | $x_+ = \frac{1}{4}(1+\bar{\varepsilon}+p)$ | $x_- = \frac{1}{4}(1+\bar{\varepsilon}-p)$ |
| | | $y_- = \frac{1}{2}(1-\bar{\varepsilon}-p)$ | $y_+ = \frac{1}{2}(1-\bar{\varepsilon}+p)$ |
| Def. | $[0,1]$ | $[0, 3-2\sqrt{2}]$ | $[0, 3-2\sqrt{2}]$ |
| $\lambda_0$ | $-\frac{1}{3}$ | $\frac{(1-3\bar{\varepsilon})p+h_1}{4}$ | $\frac{(3\bar{\varepsilon}-1)p+h_1}{4}$ |
| $\lambda_1$ | $\frac{1}{3} - 2\bar{\varepsilon}$ | $\frac{(1-3\bar{\varepsilon})p+h_2}{4}$ | $\frac{(3\bar{\varepsilon}-1)p+h_2}{4}$ |
| $\lambda_2$ | $= \lambda_1$ | $\frac{3(1-3\bar{\varepsilon})p+h_3}{4}$ | $\frac{3(3\bar{\varepsilon}-1)p+h_3}{4}$ |

$p \doteq \sqrt{\bar{\varepsilon}^2 - 6\bar{\varepsilon} + 1}$

$h_1 \doteq -3\bar{\varepsilon}^2 + 10\bar{\varepsilon} - 3, \quad h_2 \doteq -3\bar{\varepsilon}^2 + 18\bar{\varepsilon} - 3, \quad h_3 \doteq -9\bar{\varepsilon}^2 + 6\bar{\varepsilon} - 1$

### 5.2. Models with four species

As in the previous section we choose the identity matrix as reaction matrix. We will investigate three different mutation models. The first one describes a one digit string with four possible characters, the second one corresponds to a binary two digit chain, and the third one corresponds to a binary chain, where mutations with Hamming distance $d(i,j) > 1$ are forbidden.

### 5.2.1. Uniform mutation rate model

In the first case all mutation rates are equal and the mutation matrix reads

$$Q = \mathbf{Q}^{(1)} = \begin{pmatrix} 1-3\bar{\varepsilon} & \bar{\varepsilon} & \bar{\varepsilon} & \bar{\varepsilon} \\ \bar{\varepsilon} & 1-3\bar{\varepsilon} & \bar{\varepsilon} & \bar{\varepsilon} \\ \bar{\varepsilon} & \bar{\varepsilon} & 1-3\bar{\varepsilon} & \bar{\varepsilon} \\ \bar{\varepsilon} & \bar{\varepsilon} & \bar{\varepsilon} & 1-3\bar{\varepsilon} \end{pmatrix}$$

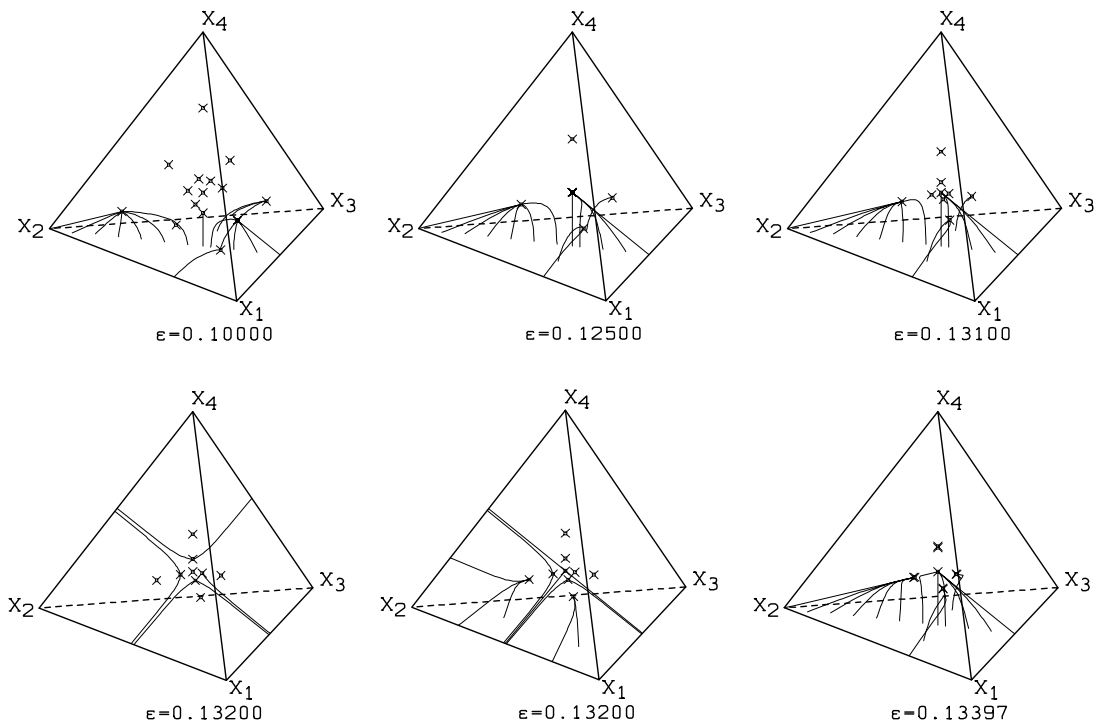

**Figure 7:** Fixed points and some trajectories for the Schlögl model with four species and the mutation matrix $W=\mathbf{Q}^{(1)}$ (uniform mutation rate model).

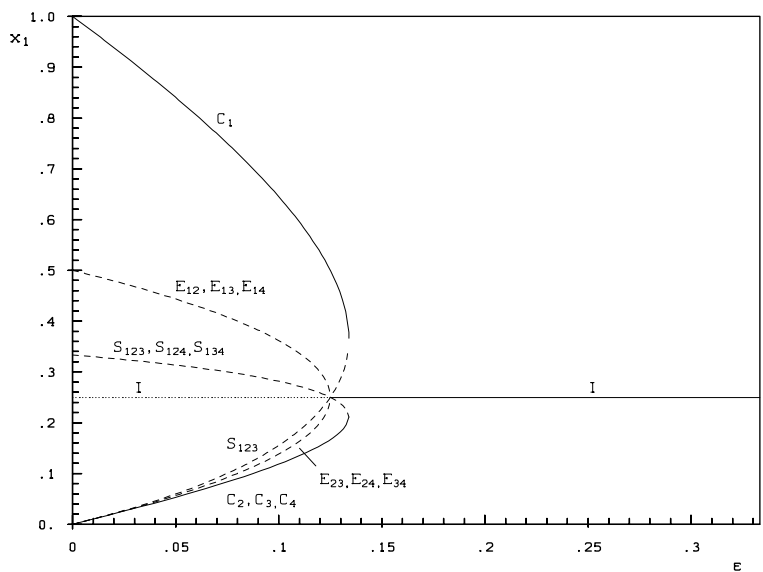

**Figure 8:** Bifurcation diagram for the Schlögl model with four species and the mutation matrix $Q=\mathbf{Q}^{(1)}$ (uniform mutation rate model). Notation as in figure 6, and $S$ denotes rest points developing from equilibria on the surface of the simplex.

**Table 3.** Fixed points and eigenvalues for the Schlögl model with four species and the mutation matrix given by $Q = \mathbf{Q}^{(1)}$ (uniform mutation rate model).

| Type | Interior | Edge | Corner | Surface |
|---|---|---|---|---|
| Coordinates | $\frac{1}{4}\mathbf{1}$ | $(x,x,y,y)$ $x = \frac{1+q_1}{4}$ $y = \frac{1-q_1}{4}$ | $(x,x,x,y)$ $x = \frac{1+2\bar{\varepsilon}-2q_2}{6}$ $y = \frac{1-2\bar{\varepsilon}+2q_2}{2}$ | $(x,x,x,y)$ $x = \frac{1+2\bar{\varepsilon}+2q_2}{6}$ $y = \frac{1-2\bar{\varepsilon}-2q_2}{2}$ |
| Def. | $[0,1]$ | $[0,\frac{1}{8}]$ | $[0, 1-\frac{\sqrt{3}}{2}]$ | $[0, 1-\frac{\sqrt{3}}{2}]$ |
| $\lambda_0$ | $-\frac{1}{4}$ | $2\bar{\varepsilon} - \frac{1}{2}$ | $\frac{2(4\bar{\varepsilon}-1)q_2 + h_1}{3}$ | $\frac{2(1-4\bar{\varepsilon})q_2 + h_1}{3}$ |
| $\lambda_1$ | $\frac{1-8\bar{\varepsilon}}{4}$ | $-\frac{1-8\bar{\varepsilon}}{2}$ | $\frac{2(4\bar{\varepsilon}-1)q_2 + h_2}{3}$ | $\frac{2(1-4\bar{\varepsilon})q_2 + h_2}{3}$ |
| $\lambda_2$ | $= \lambda_1$ | $\frac{4\bar{\varepsilon}-1}{2}q_1$ | $\frac{4(4\bar{\varepsilon}-1)q_2 + h_3}{3}$ | $\frac{4(1-4\bar{\varepsilon})q_2 + h_3}{3}$ |
| $\lambda_3$ | $= \lambda_1$ | $= -\lambda_2$ | $= \lambda_2$ | $= \lambda_2$ |

$q_1 \doteq \sqrt{1-8\bar{\varepsilon}}, \quad q_2 \doteq \sqrt{\bar{\varepsilon}^2 - 2\bar{\varepsilon} + \frac{1}{4}}$

$h_1 \doteq -8\bar{\varepsilon}^2 + 10\bar{\varepsilon} - 2, \quad h_2 \doteq -8\bar{\varepsilon}^2 + 16\bar{\varepsilon} - 2, \quad h_3 \doteq -16\bar{\varepsilon}^2 + 8\bar{\varepsilon} - 1$

In the mutation field this matrix is tantamount to the mutation matrix $\mathbf{W}^{(1)}$. Because of the high symmetry of our problem, the calculation of the fixed points simplifies considerably. There are four classes of fixed points (table 3). We have

1 central fixed point        $(\frac{1}{4}, \frac{1}{4}, \frac{1}{4}, \frac{1}{4})$

4 corner type fixed points   $(x,x,x,y)$

6 edge type fixed points     $(x,x,y,y)$

4 surface type fixed points  $(x,x,x,y)$

with all permutations of $x, y$ being allowed.

We find that bifurcations occur for $\bar{\varepsilon} = \frac{1}{8} = 0.125$ and for $\bar{\varepsilon} = 1 - \frac{\sqrt{3}}{2} \approx$

0.133975. For $\bar{\varepsilon} = 0$ the interior fixed point is a source, the corner-type equilibria are sinks, all other fixed points are saddles. When $\bar{\varepsilon}$ increases, all fixed points move into the interior of $S_n$. At $\bar{\varepsilon} = \frac{1}{8}$ the six edge-type fixed and the surface type fixed points collide with the central equilibrium. The edge type points disappear, and one eigenvalue of the surface-type points chances sign, but the equilibria remain saddles when the come out on the other side of the interior fixed point. As for the three species model, the interior equilibrium becomes stable. We have five coexisting sinks, four of which are the corner type. Their basins are separated from each other by a narrow zone with tetrahedral symmetry that form the basin of central equilibrium. For $\bar{\varepsilon} = 1 - \frac{\sqrt{3}}{2}$. the corner-type and the surface-type fixed points crash and vanish. Figure 7 gives some phase portrait for various values of $\bar{\varepsilon}$. In figure 8 we give a constraint response plot for this model.

5.2.2. The hypercube model

Individual species in this model are indentified with binary strings, for example we have $\mathbf{I}_1 = (00)$, $\mathbf{I}_2 = (01)$, $\mathbf{I}_3 = (10)$ and $\mathbf{I}_4 = (11)$ for $n = 4$, and the uniform error rate assumption is applied (since we $2^\nu$ binary strings of chain length $\nu$ hypercube mutation models exist only for dimensions that are powers of two: $n = 2, 4, 8, \ldots$).

The hypercube model looks more involved at first glance. Nevertheless, we obtained fairly simple expressions for the coordinates and the eigenvalues of all fixed points. The we write down our mutation matrix, explicitly written as

$$Q = \mathbf{Q}^{(2)} = \begin{pmatrix} q^2 & q(1-q) & (1-q)^2 & q(1-q) \\ q(1-q) & q^2 & q(1-q) & (1-q)^2 \\ (1-q)^2 & q(1-q) & q^2 & q(1-q) \\ q(1-q) & (1-q)^2 & q(1-q) & q^2 \end{pmatrix}$$

is circulant and we must have cyclic symmetry in the coordinates. In this model, however, the edges of the simplex $S_4$ are no longer equivalent. Instead, there are two classes: four edges connect corners with Hamming-distance $d(i,j) = 1$ and the remaining two edges connect corners with $d(i,j) = 2$. We refer to them as

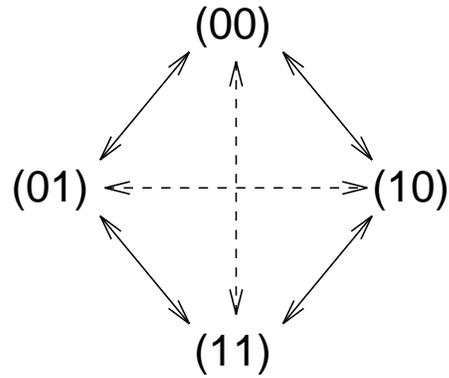

**Figure 9:** The hypercube mutation rate model for four species. Full arrows ($<$—$>$) refer to single point mutations, and dotted arrows ($<\cdots>$) indicate simultaneous changes of two digits at a time.

*direct* and *indirect* edges, respectively. There are also two different edge-type fixed points, and thus we are dealing with five classes of fixed points (table 4).

| | |
|---|---|
| 1 central fixed point | $(\frac{1}{4}, \frac{1}{4}, \frac{1}{4}, \frac{1}{4})$ |
| 4 corner type fixed points | $(x, y, x, z)$ |
| 4 direct edge type fixed points | $(x, x, y, y)$ |
| 2 indirect edge type fixed points | $(x, y, x, y)$ |
| 4 surface type fixed points | $(x, y, x, z)$ |

For the mutation matrix $\mathbf{Q}^{(2)}$ there exist four different bifurcation points:

$$q_1 = \frac{\sqrt{3}}{2} \approx 0.866, \qquad q_2 = \frac{1}{2}(1 + \frac{1}{\sqrt{2}}) \approx 0.8536$$
$$q_3 = \frac{3}{4}, \qquad q_4 = \frac{1}{2}(1 - \frac{1}{\sqrt{2}}) \approx 0.1464.$$

For $q = 1$ we have the unperturbed Schlögl model. As $q$ decreases, all fixed points move inwards. The central equilibrium is a source, the corner type fixed points are sinks, and we have 10 saddle points in $S_n$. At $q = q_1$ two pairs of surface type saddles collide with indirect edge type fixed points, one eigenvalue of which changes sign, and disappear. The indirect edge type equilibria remain saddles until $q = q_2$ where they vanish crashing into the central fixed point. At this critical value of $q$ the central equilibrium becomes a saddle, and remains so till

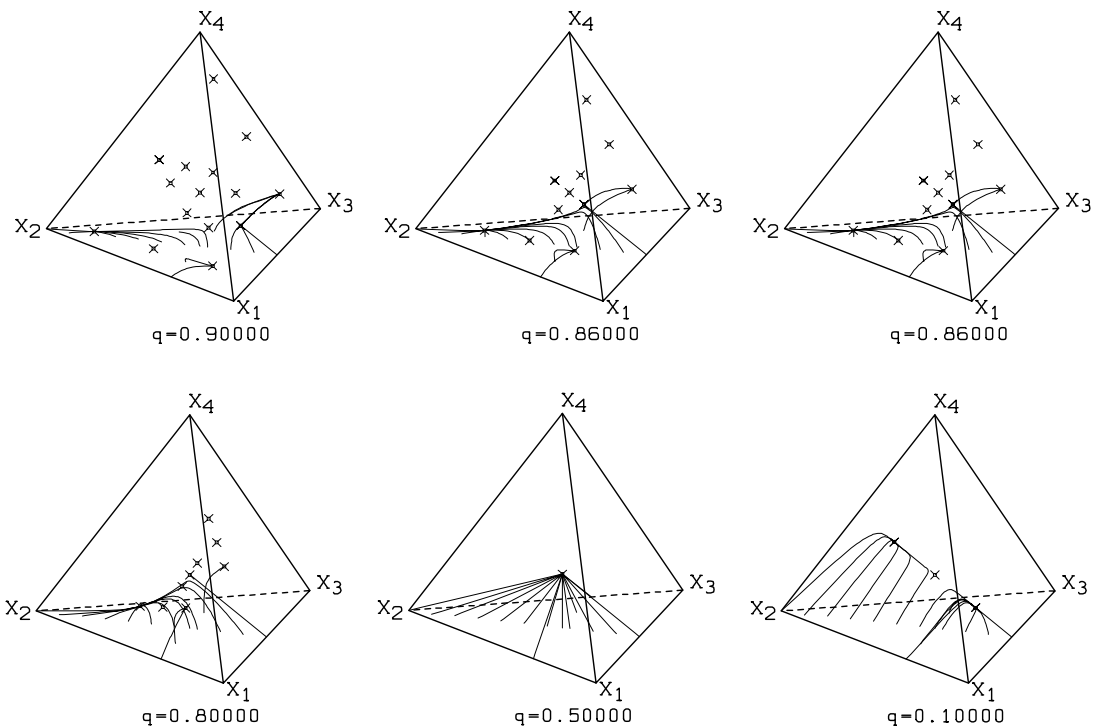

**Figure 10:** Fixed points and trajectories for the four species Schlögl model with mutation matrix $Q=\mathbf{Q}^{(2)}$ (hypercube model).

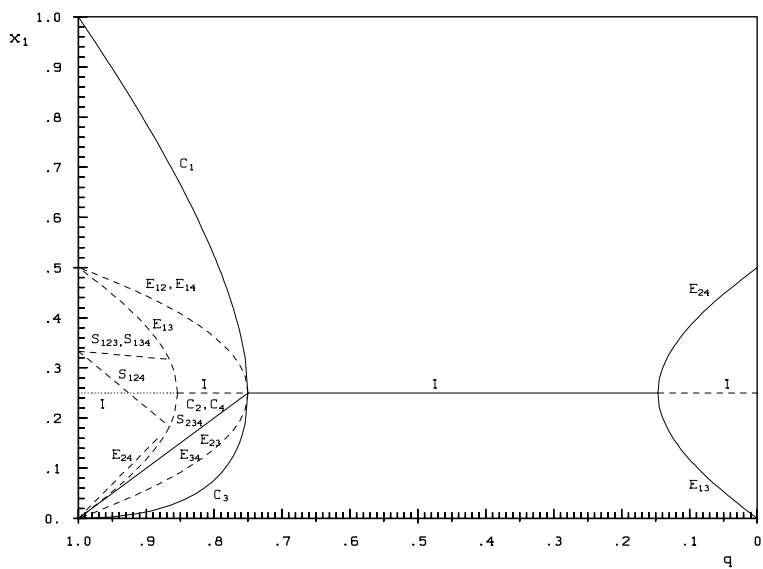

**Figure 11:** Bifurcation diagram for the four species Schlögl model with mutation matrix $Q=\mathbf{Q}^{(2)}$ (hypercube model).

**Table 4.** Fixed points and eigenvalues for the four-species Schlögl model with the mutation matrix given by $\mathbf{Q}^{(2)(2)}$ (hypercube model).

| Type | Interior | Corner | direct Edge | indirect Edge | Surface |
|---|---|---|---|---|---|
| Coord. | $\frac{1}{4}\mathbf{1}$ | $(x,y,x,z)$ $x=1-q$ $y=\frac{f+q_1}{2}$ $z=\frac{f-q_1}{2}$ | $(x,x,y,y)$ $x=\frac{1+q_1}{4}$ $y=\frac{1-q_1}{4}$ | $(x,y,x,y)$ $x=\frac{1+q_2}{4}$ $y=\frac{1-q_2}{4}$ | $(x,y,x,z)$ $x=\frac{q}{2q+1}$ $y=\frac{1+q_3}{2(2q+1)}$ $z=\frac{1-q_3}{2(2q+1)}$ |
| Def. | $[0,1]$ | $[\frac{3}{4},1]$ | $[\frac{3}{4},1]$ | $[0,\frac{1}{2}-\frac{1}{\sqrt{8}}]\cup$ $\cup[\frac{1}{2}+\frac{1}{\sqrt{8}},1]$ | $[\frac{\sqrt{3}}{2},1]$ |
| $\lambda_0$ | $-\frac{1}{4}$ | $-f^2$ | $-\frac{f}{2}$ | $-\frac{f^2}{2}$ | $-\frac{f}{2q+1}$ |
| $\lambda_1$ | $\frac{4q-3}{4}$ | $-f(4q-3)$ | $-\frac{4q-3}{2}$ | $-\frac{8q^2-8q+1}{2}$ | $\frac{f^2}{2q+1}$ |
| $\lambda_2$ | $=\lambda_1$ | $=\lambda_1$ | $\frac{(4q-3)f}{2}$ | $-\frac{f}{2}(f-(1+q_2))$ | $\frac{2q^3-q^2-3q+2+q_4}{2q+1}>0$ |
| $\lambda_3$ | $\frac{8q^2-8q+1}{4}$ | $1-2q^2$ | $-\frac{f^2}{2}$ | $-\frac{f}{2}(f-(1-q_2))$ | $\frac{2q^3-q^2-3q+2-q_4}{2q+1}<0$ |

$f \doteq 2q-1, \quad q_1 \doteq \sqrt{4q-3}, \quad q_2 \doteq \sqrt{8q^2-8q+1}, \quad q_3 \doteq \sqrt{4q^2-3}$

$q_4 \doteq \sqrt{4q^6+12q^5-19q^4-6q^3+15q^2-6q+1}$

$q = q_3$. Here the corner type and the direct edge type fixed points disappear by colliding with the central equilibrium which becomes stable. Between $q_3$ and $q_4$ the central equilibrium is stable and unique. At $q = q_4$ it loses stability again and becomes a saddle; two indirect edge type stable fixed points split off the interior equilibrium and stay stable until $q = 0$. We illustrate this behaviour in figure 9, a constraint response plot for the binary chain model is given in figure 10.

5.2.3. The single point mutation model

In the third and last model mutations are restricted to single point mutations. It is a good approximaion for small error rates. The mutation matrix then reads

$$Q = \mathbf{Q}^{(3)} = \begin{pmatrix} 1-2w & w & 0 & w \\ w & 1-2w & w & 0 \\ 0 & w & 1-2w & w \\ w & 0 & w & 1-2w \end{pmatrix}$$

where we use the parameter $w \doteq \frac{3}{2}\bar{\varepsilon}$ instead of $\bar{\varepsilon}$ for convenience of calculation. The behaviour is very similar to the previous system, with the exception, that the central equilibrium does not become unstable for very low replication accuracy. As for the previous model, matrix $\mathbf{Q}^{(2)}$ we expect and find five classes of fixed points:

| | |
|---|---|
| 1 central fixed point | $(\frac{1}{4}, \frac{1}{4}, \frac{1}{4}, \frac{1}{4})$ |
| 4 corner type fixed points | $(x, y, x, z)$ |
| 4 direct edge type fixed points | $(x, x, y, y)$ |
| 2 indirect edge type fixed points | $(x, y, x, y)$ |
| 4 surface type fixed points | $(x, y, x, z)$ |

The results for the single point mutation model are summarized in table 5.

For $w = 0$ we start with the same situation as above. As the mutation increases, all fixed points move inwards. The first bifurcation occurs at

$$\alpha_0 \doteq \frac{\sqrt[3]{9\sqrt{11} - 17\sqrt{3}}^2 + 4\sqrt[6]{3} \cdot \sqrt[3]{9\sqrt{11} - 17\sqrt{3}} - 2\sqrt[3]{3}}{12\sqrt[6]{3} \cdot \sqrt[3]{9\sqrt{11} - 17\sqrt{3}}} \approx 0.114078$$

where the surface type saddles collide with the indirect edge type saddles and disappear. We remark, that $\alpha_0$ is the unique real zero of

$$32x^3 - 32x^2 + 12x - 1 = 0$$

The indirect edge type saddles change the sign of one eigenvalue and remain saddles; the collide with the central source at $w = \frac{1}{8}$, disappearing and leaving the

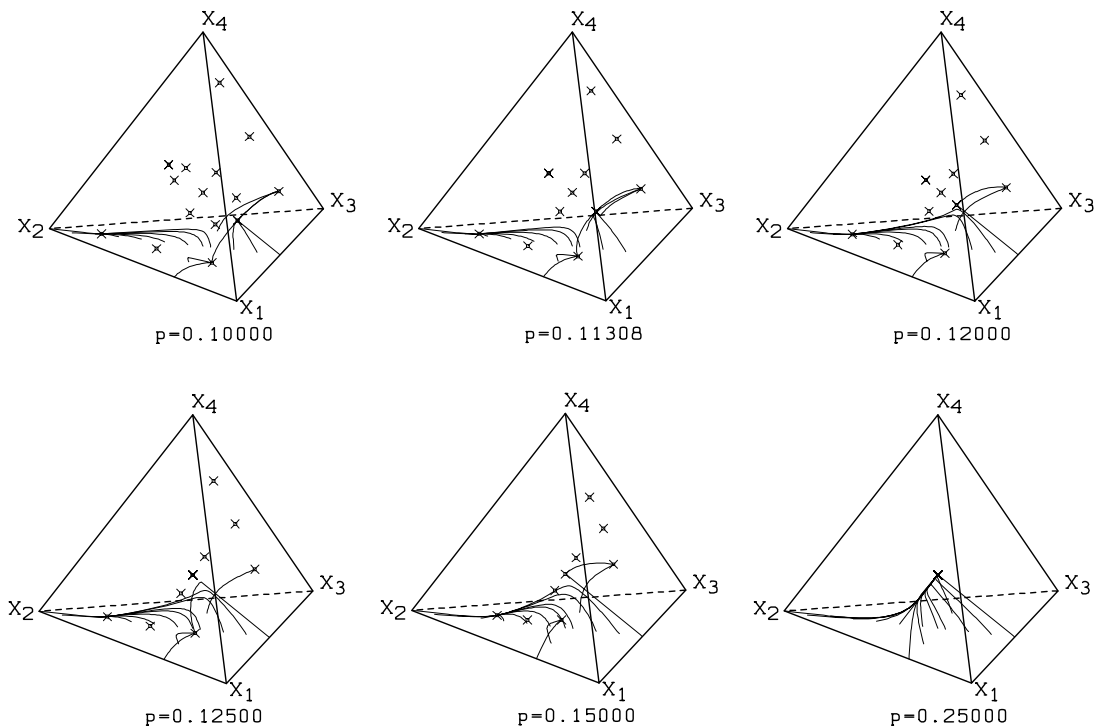

**Figure 12:** Equilibria and trajectories for the four species Schlögl model with mutation matrix $Q=\mathbf{Q}^{(3)}$ (single point mutation model).

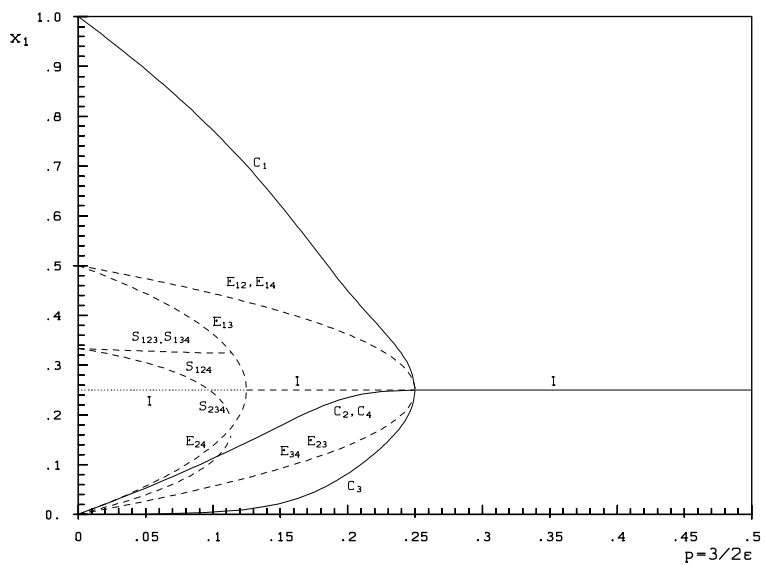

**Figure 13:** Bifurcation diagram for the four species Schlögl model with mutation matrix $Q=\mathbf{Q}^{(3)}$ (single point mutation model).

**Table 5.** Fixed points and eigenvalues for the four-species Schlögl model with mutation matrix $Q=\mathbf{Q}^{(3)}$ (single point mutation model).

| Type | Interior | Corner | direct Edge | indirect Edge | Surface |
|---|---|---|---|---|---|
| Coord. | $\frac{1}{4}\mathbf{1}$ | $(x,y,x,z)$ $x=\frac{4w^2-1+\sqrt{f}}{2(8w-3)}$ $y=\frac{v-\sqrt{f}+\sqrt{g+h\sqrt{f}}}{2(8w-3)}$ $z=\frac{v-\sqrt{f}-\sqrt{g+h\sqrt{f}}}{2(8w-3)}$ | $(x,x,y,y)$ $x=\frac{1+q_1}{4}$ $y=\frac{1-q_1}{4}$ | $(x,y,x,y)$ $x=\frac{1+q_2}{4}$ $y=\frac{1-q_2}{4}$ | $(x,y,x,z)$ $x=\frac{4w^2-1-\sqrt{f}}{2(8w-3)}$ $y=\frac{v+\sqrt{f}+\sqrt{g-h\sqrt{f}}}{2(8w-3)}$ $z=\frac{v+\sqrt{f}-\sqrt{g-h\sqrt{f}}}{2(8w-3)}$ |
| Def. | $[0,1]$ | $[0,\frac{1}{4}]$ | $[0,\frac{1}{4}]$ | $[0,\frac{1}{8}]$ | $[0,\alpha_0]$ |
| $\lambda_0$ | $-\frac{1}{4}$ | $<0$ | $\frac{2w-1}{2}$ | $\frac{4w-1}{2}$ | $<0$ |
| $\lambda_1$ | $\frac{1-8w}{4}$ | $<0$ | $\frac{4w-1}{2}$ | $\frac{8w-1}{2}$ | $<0$ |
| $\lambda_2$ | $\frac{1-4w}{2}$ | $\leq 0$ | $-\frac{w+\sqrt{q_3}}{2}$ | $w+(w-\frac{1}{2})q_2$ | $>0$ |
| $\lambda_3$ | $=\lambda_2$ | $\leq 0$ | $-\frac{w-\sqrt{q_3}}{2}$ | $w-(w-\frac{1}{2})q_2$ | $>0$ |

$q_1 \doteq \sqrt{1-4w}, \quad q_2 \doteq \sqrt{1-8w}, \quad q_3 \doteq -32w^3+33w^2-10w+1$

$f \doteq 16w^4-64w^3+48w^2-12w+1, \quad h \doteq 16w^2-12w+4$

$g \doteq 64w^4-176w^3+144w^2-48w+5, \quad v \doteq -4w^2+8w-2$

$\alpha_0 \approx 0.1140777468$

interior equilibrium as saddle point. For $w=\frac{1}{4}$ the direct edge type saddles and the corner type sinks collide with the interior equilibrium and annihilate. The central fixed point is left unique and stable. We find this model with mutation matrix $\mathbf{Q}^{(3)}$ to have only one qualitative difference with the model with the $Q$ matrix: The central equilibrium remains stable till the end of physical domain, and the indirect edge type point do not reappear for large values of $w$ (or $\overline{\varepsilon}$). We show some phase portrait in figure 11 and give a constraint response plot for this model in figure 12.

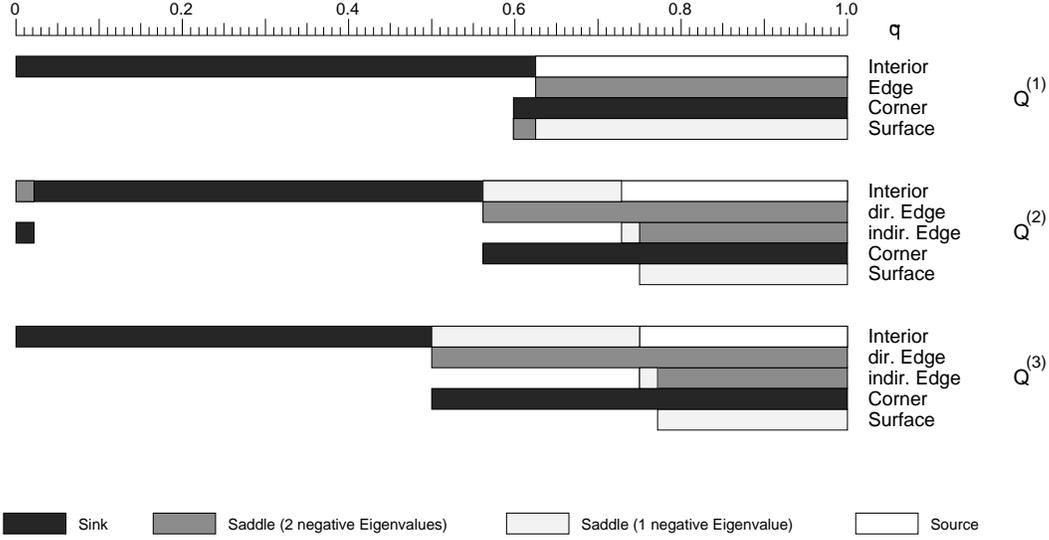

**Figure 14:** Comparison of the stability of the fixed points for different mutation matrices. All mutation parameters have been converted to the mean replication accuracy $\tilde{q}$ ($\mathbf{Q}^{(1)}$: uniform error rates; $\mathbf{Q}^{(2)}$: hypercube model; $\mathbf{Q}^{(3)}$ single point mutations).

5.2.4. Comparison of the three models

In order to compare the three mutation models the coordinates and eigenvalues of all fixed points were expressed as functions of the same parameter, the mean replication accuracy $\tilde{q} = \frac{1}{n} \sum_{k=1}^{n} q_{kk}$:

(1)  uniform mutation rate model:   $\tilde{q} = 1 - (n-1)\overline{\varepsilon}$

(2)  hypercube model:   $\tilde{q} = q^{\nu} = q^{\operatorname{ld} n}$

(3)  single point mutation model:   $\tilde{q} = q^{\nu} = 1 - \nu \cdot w = 1 - (\operatorname{ld} n) \cdot w$

The results are shown in figure 14. In all three models we observe a bifurcation at which a stable rest point jumps from a corner (representing the species with the highest fitness value) towards the center. This bifurcation is tantamount to an error threshold of replication which will be discussed in the final section 6. We observe two interesting differences:

(1) In the uniform mutation rate model this bifurcation is subcritical, wherease it is supercritical in the other two cases. Thus the uniform mutation rate model can lead to hysteresis phenomena.

(2) In the hypercube model the rest point in the center becomes again unstable at very small values of $\tilde{q}$. Then the stable fixed point jumps onto the indirect edge. It corresponds to inaccurate complemetary replication (Swetina and Schuster, 1982; Schuster and Swetina, 1988; Stadler, 1991). In the limit $\tilde{q} \to 0$ we have precise complementary replication: $A + \mathbf{I}_\oplus \to \mathbf{I}_\ominus + \mathbf{I}_\oplus$ and $A + \mathbf{I}_\ominus \to \mathbf{I}_\oplus + \mathbf{I}_\ominus$.

Apart from the stable rest point on the indirect edge (hypercube model) all rest points on edges and surfaces are unstable.

### 5.3. The central equilibrium in high-dimensional systems

The Jacobian for model (23) at the inner fixed point $\mathbf{c}$ reads

$$D = \frac{1}{n}\left[(2+n\alpha)Q - (1+n\alpha)E - \frac{2}{n}\mathbf{I}\right]. \qquad (24)$$

Obviously, $D$ is circulant if and only if $Q$ is circulant.

**Theorem 5.1.** The central equilibrium $\mathbf{c}$ of Schlögl's model, equation (23), with mutation matrix $\mathbf{Q}^{(1)}$ is a sink for $\overline{\varepsilon} > \overline{\varepsilon}_C$ and a source for $\overline{\varepsilon} < \overline{\varepsilon}_C$, where

$$\overline{\varepsilon}_C \doteq \frac{1}{n} \cdot \frac{1}{2+n\alpha}.$$

**Proof.** $W$ is circulant and $q_{ik} = \overline{\varepsilon}$ for $i \neq k$ and $q_{kk} = 1 - (n-1)\overline{\varepsilon}$ for diagonal elements of $Q$. Hence equation (24) becomes

$$d_{kl} = \frac{1}{n}\left[(2+n\alpha)\overline{\varepsilon} - \frac{2}{n}\right] \quad \text{for } k \neq l \quad \text{and}$$

$$d_{kk} = \frac{1}{n}\left[1 - (n-1)(2-n\alpha)\overline{\varepsilon} - \frac{2}{n}\right].$$

$D$ is circulant due to Lemma 5.1 and therefore has the eigenvalues

$$\mu_0 = d_{11} + (n-1)d_{12} = -\frac{1}{n},$$

$$\mu_1 = d_{11} - d_{12} = \frac{1}{n}[1 - n(2+n\alpha)\bar{\varepsilon}].$$

It is easy to verify that $D \cdot \mathbf{1} = -\frac{1}{n}\mathbf{1}$, i.e., $\mu_0$ belongs to the unphysical external direction, and the dynamical behaviour is determined by the $(n-1)$-fold degenerate eigenvalue $\mu_1$ alone. ∎

**Lemma 5.2.** Let $Q$ be symmetric. Then each eigenvector of $Q$ is also an eigenvector of $D$. The eigenvalues of $D$ are

$$\mu_k = \frac{1}{n}\Big[(2+n\alpha)\phi_k - (1+n\alpha) - 2\delta_{k,0}\Big]$$

where $\phi_k$ are the eigenvalues of $Q$.

**Proof.** Since $Q$ is stochastic and symmetric $\mathbf{1}$ is an eigenvector, and all other eigenvectors $\mathbf{z}_k$ can be chosen to be orthogonal to $\mathbf{1}$. Let $\mathbf{I}$ denote the matrix with all entries equal to 1. Multiplying the Jacobian, equation (24), with the eigenvectors $\mathbf{z}_k$ of $Q$ yields

$$D \cdot \mathbf{1} = \frac{1}{n}\Big[(2+n\alpha)P \cdot \mathbf{1} - (1+n\alpha) \cdot \mathbf{1} - \frac{2}{n}\mathbf{I} \cdot \mathbf{1}\Big] =$$
$$= \frac{1}{n}\Big[(2+n\alpha)\phi_0 - (1+n\alpha) - \frac{2}{n}\cdot n\Big] \cdot \mathbf{1} \doteq \mu_0 \cdot \mathbf{1}$$
$$D \cdot \mathbf{z}_k = \frac{1}{n}\Big[(2+n\alpha)P \cdot \mathbf{z}_k - (1+n\alpha) \cdot \mathbf{z}_k - \frac{2}{n}\mathbf{I} \cdot \mathbf{z}_k\Big] =$$
$$= \frac{1}{n}\Big[(2+n\alpha)\phi_k - (1+n\alpha) - 0\Big] \cdot \mathbf{z}_k \doteq \mu_k \cdot \mathbf{z}_k$$

∎

We can now determine the stability of the central rest point $\mathbf{c}$ of Schlögl's model with mutation terms defined by the uniform error model for arbitrary chain length $\nu$.

**Theorem 5.2.** For the Schlögl model with mutation matrix $\mathbf{Q}^{(2)}$ holds:

(i) The central fixed point **c** is a source if

$$q > \frac{1}{2}\Big(1 + \sqrt[\nu]{\frac{1}{2}\frac{1+2^\nu\alpha}{1+2^{\nu-1}\alpha}}\Big) = q_\nu^+.$$

(ii) The central fixed point **c** is a sink if

$$\frac{1}{2}\Big(1 - \sqrt[2]{\frac{1}{2}\frac{1+2^\nu\alpha}{1+2^{\nu-1}\alpha}}\Big) < q < \frac{1}{2}\Big(1 + \frac{1}{2}\frac{1+2^\nu\alpha}{1+2^{\nu-1}\alpha}\Big).$$

For the Schlögl model with mutation matrix $\mathbf{Q}^{(3)}$ holds:

(i) The central fixed point **c** is a source if

$$q > \frac{1}{1 + \frac{1}{\nu(2(2+2^\nu\alpha)-1)}} = \tilde{q}_\nu.$$

(ii) The central fixed point **c** is a sink if

$$0 \leq q < \frac{2(2+2^\nu\alpha) - \nu}{2(2+2^\nu\alpha) - \nu + 1} \quad \text{and}$$

$$\alpha > \frac{\nu - 4}{2^{\nu+1}}$$

The second condition is always fulfilled if $\nu \leq 4$.

**Proof.** The eigenvalues for the mutation models $\mathbf{Q}^{(2)}$ and $\mathbf{Q}^{(3)}$, respectively, are easily calculated from equation (24). We find

$$\begin{aligned}
\mu_0(\nu) &= \tilde{\mu}_0(\nu) = -\frac{1}{n} = -2^{-\nu} \\
\mu_j(\nu) &= \frac{1}{2^\nu} \cdot \big[(2+2^\nu\alpha)(2q-1)^j - (1+2^\nu\alpha)\big] \\
\tilde{\mu}_j(\nu) &= \frac{1}{2^\nu} \cdot \Big[(2+2^\nu\alpha)\frac{q + (\nu-2j)(1-q)}{q+\nu(1-q)} - (1+2^\nu\alpha)\Big]
\end{aligned} \quad (25)$$

The eigenvalues $\mu_j(\nu)$, $\tilde{\mu}_j$ become zero for

$$\begin{aligned}
q_j^\pm &= \frac{1}{2}\Big(1 \pm \sqrt[j]{\frac{1}{2}\cdot\frac{1+2^\nu\alpha}{1+2^{\nu-1}\alpha}}\Big) \\
\tilde{q}_j &= \frac{2j(2+2^\nu\alpha) - \nu}{2j(2+2^\nu\alpha) - (\nu-1)}
\end{aligned}$$

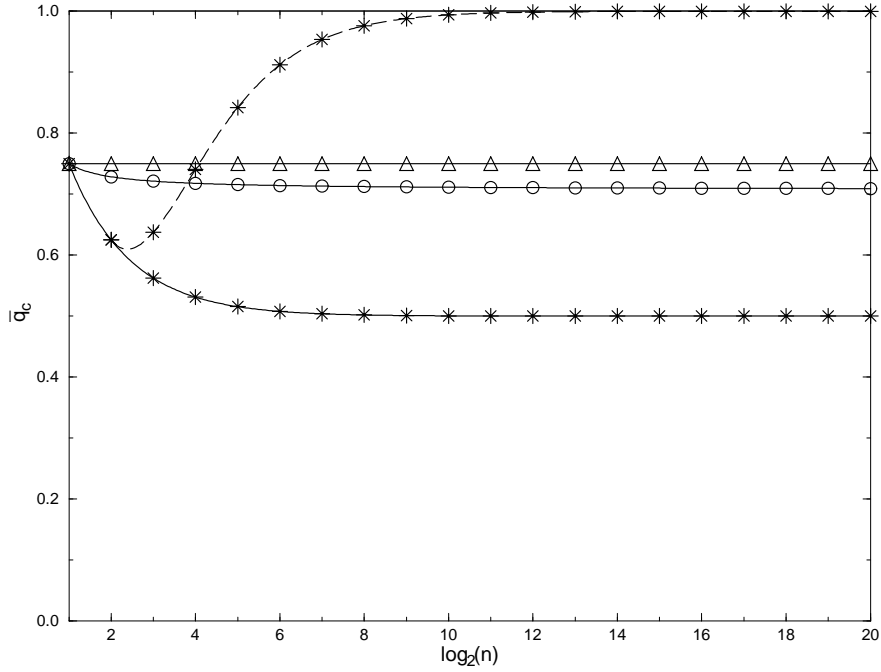

**Figure 15:** Critical mean replication accuracy per replication round $\bar{q}_c$ for different models and mutation matrices (∗ for mutation matrix $\mathbf{Q}^{(1)}$, ○ for $\mathbf{Q}^{(2)}$, and △ for $\mathbf{Q}^{(3)}$) depending on the number of species. Solid lines denote Schlögel's Model and the dashed line stands for the Hypercycle.

where $q_j^-$ is a zero of $\mu_j(\nu)$ if and only if $j$ is *even* and $\tilde{q}_j$ is a zero in the physically meaningful domain, i.e. $\tilde{q}_j \in [0,1]$, if and only if $j \geq \frac{\nu}{2(2+2^\nu \alpha)}$. ∎

The parameter $q$ ist related to the mean replication accuracy $\bar{q}$ per replication round. For the full model $\mathbf{Q}^{(2)}$, the simplified model $\mathbf{Q}^{(3)}$, and the mutation matrix $\mathbf{Q}^{(1)}$ we find

$$\bar{q}[\mathbf{Q}^{(2)}] = q^\nu, \qquad \bar{q}[\mathbf{Q}^{(3)}] = \frac{q}{q + \nu(1-q)}, \qquad \text{and} \qquad \bar{q}[\mathbf{Q}^{(1)}] = q - (n-1)\bar{\varepsilon},$$

respectively. We are interested in the first bifurcation at the interior fixed point when we switch on the mutation, $\bar{q}_c$. Since we have used different mutation parameters for convenience of calculation we transform them to $\bar{q}$ in order to compare them. For simplicity we set $\alpha = 0$ and $n = 2^\nu$ since the matrices $\mathbf{Q}^{(2)}$ and $\mathbf{Q}^{(3)}$ can be defined only for powers of 2. We find

$$\bar{q}_c[\mathbf{Q}^{(1)}] = \frac{2^\nu + 1}{2^{\nu+1}},$$

$$\bar{q}_c[\mathbf{Q}^{(2)}] = [\frac{1}{2}(1 + \sqrt[\nu]{\frac{1}{2}})]^\nu,$$

$$\bar{q}_c[\mathbf{Q}^{(3)}] = \frac{3}{4}.$$

For the hypercycle model with $\mathbf{Q}^{(1)}$ we find

$$\bar{q}_c[\mathbf{Q}^{(1)}] = 1 - (1 - \frac{1}{2^\nu})\frac{\cos\frac{\pi}{2}(1 - 2^{2-\nu})}{1 + \cos\frac{\pi}{2}(1 - 2^{2-\nu})}$$

We find that the parameter range in which the interior fixed point of the mutating system exhibits the same dynamical behaviour as in the pure replicator equation depends strongly on both the reaction matrix $A$ and the mutation matrix $Q$. Note that the three different mutation matrices reduce to the unique matrix $\mathbf{Q}^{(2)}(1)$ for $n = 2$, i.e, for $\nu = 1$. The values of $\bar{q}_c$ are compares in figure 14.

## 6. Conclusions

The results obtained for "catalyzed autocatalysis" (second order autocatalysis: equ.1) with different model mutation models share the existence of an error threshold phenomenon with the simple replication-mutation system with frequency independent replication rates (first order autocatalysis: Eigen, 1971; Eigen and Schuster, 1977; Eigen *et al.*, 1988 and 1989). In this simple case the system converges towards a unique and asymptotically stable stationary state provided there is no degeneracy of fitness values (i.e. the fitness of the fittest species called master species is larger than that of any other species) and the mutation matrix $Q$ is positive definite. The mutant distribution is obtained as the largest eigenvector $\zeta_1$ of the matrix $Q \cdot (\vec{a} \cdot E)$ where $\vec{a} = (a_1, a_2, \ldots, a_n)$, the vector of replication rate constants. From Frobenius theorem follows that all components of $\zeta_1$ are strictly positive. The eigenvector $\zeta_1$ when computed as a function of the error rate $\overline{p}$ shows a threshold phenomenon at some critical $\overline{p}$-value which is reminiscent of cooperative transitions. At low values of $\overline{p}$ the mutant distribution is centered around a master species (or master sequence if polynucleotides are considered). In terms of phase portraits this means that the unique stable fixed point $P$ is located near a corner of the simplex $S_n$. As a function of $\overline{p}$ the equilibrium point moves slowly towards the center **c**. In the transition region around the critical error rate which is tantamount to the error threshold $P$ jumps suddenly towards **c** and reaches the center asymptotically when replication becomes completely random ($\overline{p} = \frac{1}{2}$ in the hypecube model): $\lim_{\overline{p} \to 1/2} P(\overline{p}) = \mathbf{c}$. The longer the sequence $(\nu)$, or the larger the number of species $(n = 2^\nu)$ is, the narrower becomes the transition zone, the sharper is the threshold. Depending on the vector of replication rate constants $\vec{a}$ the width of the transition region may remain finite in the limit of large numbers of species $(\lim n \to \infty)$ which is characteristic for a cooperative transition, or it may shrink to a transition point. Then we are dealing with a (higher order) phase transition (Leuthäusser, 1987).

The multi-dimensional Schlögl model (Schuster, 1986) is the higher order analogue to the simple selection case (with frequency independent replication rates). Instead of a unique stable stationary state we find $n$ equlibria, one at each corner of the simplex $S_n$. Rate constants determine the sizes of the basins of attraction. Mutation causes the stable equilibria to move into the interior of $S_n$. At some critical error rate we observe a bifurcation (or several bifurcations) which make the central rest point to the unique stable steady state of the system. This is straightforward to visualize since the central rest point is the unique fixed point of the mutation matrix. Here, the error threshold resembles a first order phase transition. This result was obtained for all mutation models, although some differences are observed in detail: in the uniform mutation rate model the bifurcation is subcritical and the system shows hysteresis accordingly, whereas we find supercritical single jumps towards the center in the other two cases.

In general we observe that the onset of mutation simplifies replicator dynamics: multiple stable equilibria in the Schlögl model change into a single stable restpoint at sufficiently high error rates, oscillations in the mutation-free hypercycle model are replaced by the central equilibrium, and the chaotic attractor of a replicator equation with four species is turned stepwise into simpler dynamics if mutation terms are introduced.

### Acknowledgements

This work was supported financially by the Austrian *Fonds zur Förderung der Wissenschaftlichen Forschung* (Projects P-4506, P-5286 and P-6864) as well as by the *Stiftung Volkwagenwerk*, Germany.

# Contents